\documentclass[%
preprint,
nobibnotes,
 amsmath,amssymb,
aps,
prb,
floatfix,
]{revtex4-1}
\usepackage{graphicx}
\usepackage{dcolumn}
\usepackage{bm}
\usepackage{latexsym,indentfirst}
\usepackage{graphicx}
\usepackage[caption=false]{subfig}
\usepackage{epstopdf}
\usepackage{epsfig}
\usepackage{slashbox}
\usepackage{multirow}
\usepackage[colorlinks,linkcolor=blue,anchorcolor=blue,citecolor=blue]{hyperref}

\begin{document}

\title{A Fokker-Planck reaction model for the epitaxial growth and shape transition of quantum dots}

\author{Chaozhen Wei}
\affiliation{Department of Mathematics, University at Buffalo, State University of New York, Buffalo, NY 14260-2900, USA}
\author{Brian J. Spencer}
\affiliation{Department of Mathematics, University at Buffalo, State University of New York, Buffalo, NY 14260-2900, USA}


\date{\today}

\begin{abstract}
We construct a Fokker-Planck reaction model to investigate the dynamics of the coupled epitaxial growth and shape transition process of an array of quantum dots. The Fokker-Planck reaction model is based on a coupled system of Fokker-Planck equations wherein the distribution of each island type is governed by its own Fokker-Planck equation for growth, with reaction terms describing the shape transitions between islands of different types including asymmetric shapes. The reaction terms for the shape transitions depend on the island size and are determined from explicit calculations of the lowest-barrier pathway for each shape transition. This mean-field model enables us to consider the kinetics of asymmetric shape transitions and study the evolution of island shape distributions during the coupled growth and transition process. Through numerical simulations over a range of growth parameters, we find multimodal and unimodal evolution modes of the shape distribution of island arrays, which depend on the external deposition flux rate and temperature rather than the shape transition rate. However, the shape transition rate governs the kinetics of shape transitions and determines the fraction of islands that form via asymmetric states, which has implications for the development of asymmetric composition profiles within alloy islands.
\end{abstract}

\maketitle

\section{Introduction} \label{section:intro}
Quantum dots (QDs) are one kind of important semiconductor nanostructure which has prospective applications in microelectronic and optoelectronic devices due to the quantum confinement effect and the possibility of tuning their properties through controlling their size and shape during the preparation. The self-assembly of strained islands during epitaxial growth is a promising method for fabricating QDs which has attracted intense interest in the past decades. \cite{Aqua2013Review,Li2014Review} However, it is still a significant challenge to create dense arrays of uniform QDs with desired shape, size and composition because islands of different types and sizes may coexist during the growth process \cite{Ross1998Coarsening,ross1999transition,Rastelli2005} and the distribution of composition within islands can be complex. \cite{Spencer2001Morphological,alloiedprepyramid,Tersoff2007Coarsening,Han2010composition,Shenoy2011} Thus a key problem in developing self-assembled nanostructures is to understand the QDs growth and shape transition process, and hence be able to control the island size, shape and material composition by manipulating the material properties and growth conditions.\par
In the Stranski-Krastanov growth mode of thin films, self-assembled QDs, also termed ``islands", form spontaneously as a mechanism to relieve the misfit strain between the film and substrate. Moreover, as the islands grow in size they undergo a sequence of shape transitions where in general the island shape has a strong size dependence. In the prototypical Ge/Si system, small islands have a pyramidal shape with a rectangular base, called ``pyramid"; large islands have a multifaceted shape which has multiple steep facets near the foot of the pyramid, called ``dome". Because of the complexity of the epitaxial growth process, a large amount of research has been done to study the size and shape distribution of QDs in an array of islands during growth in order to get control on the QD distribution in production. \cite{Medeiros1998,Rudd2003,Rudd2007,Ross1998Coarsening,Floro2000,Jesson2004Suppression,jesson2004metastability,Vine2005,Munt2007,Nevalainen2007Size,Pirkkalainen2008twocomputational,lam2010kinetic,Aqua2013simulation,Williams2000,Meixner2001,Rastelli2005} \par
One of the most important phenomenon observed in experiments is the coexistence of different island shapes and the multimodal size distribution of islands. \cite{Medeiros1998,Ross1998Coarsening,Rastelli2005} During growth, pyramids and domes can co-exist not only at two different characteristic sizes within an array of islands resulting in a bimodal size distribution, but also can co-exist at the same size where the shape transition occurs. Regarding the coexistence of pyramid and dome and their bimodal distribution, two competing views (equilibrium thermodynamics versus kinetics) are presented and compared. \cite{Williams2000,Meixner2001,Rastelli2005} The thermodynamic model suggests that the coexistence and the bimodal distribution of pyramid and dome reflects two minima in the island energy at two different volumes. \cite{Medeiros1998,Rudd2003,Rudd2007} On the other hand, the kinetic model considers the evolution of the island distribution as a result of a kinetically-limited process. \cite{Ross1998Coarsening,Floro2000,Jesson2004Suppression,jesson2004metastability,Vine2005,Munt2007,Nevalainen2007Size,Pirkkalainen2008twocomputational,lam2010kinetic,Aqua2013simulation}  In this kinetic process, a coarsening process is observed where islands larger than a critical size are growing while smaller islands are shrinking, which results in the bimodal distribution without need of energy minima. \cite{Rastelli2005} This coarsening process can be explained by a mean-field kinetic model \cite{Ross1998Coarsening,Vine2005} in which the shape transition between pyramid and dome introduces a jump in chemical potential so that large domes (with lower chemical potential) grow rapidly while small pyramids (with higher chemical potential) grow slowly or even shrink. \par
Another important feature of island growth is shape transitions involving asymmetric island shapes during the kinetic process. These asymmetric shapes are observed in experiment, \cite{ross1999transition} and investigated in simulation \cite{lam2010kinetic} and theory. \cite{Spencer2013,Wei20160262} The asymmetric island shape can be a metastable state during the shape transition and its stability is dependent on the energy barrier along the transition path. Since the asymmetric transition has different energy barriers from a presumed symmetric transition, it will affect the kinetics of shape transitions and the shape distribution of islands. Moreover, the asymmetric shape transition phenomenon is of particular interest because it could cause asymmetric composition profiles within alloyed islands that will affect the properties of nanostructured devices. Thus we develop a model to take into account the kinetic process of shape transitions during the epitaxial growth process. \par
In this paper, we investigate the dynamics of the growth-transition process of an array of islands by using a kinetic Fokker-Planck reaction (FPR) model, which is based on a system of Fokker-Planck equations coupled with reaction terms that describe the shape transitions between different island types. Moreover, we utilize detailed calculations of the energy of islands as a function of island shape and size from Ref.~\onlinecite{Wei20160262} including the lowest-barrier shape transition path, which often involves asymmetric islands. We use the transition barriers to determine the reaction coefficients between island types, which enables us to study the effect of asymmetric transition states on how the shape and size distribution of islands evolves during film growth. \par
The rest of our paper is organized as follows. In Sec.~\ref{section:model}, we will introduce our energy model for 2D faceted islands and construct the FPR model for describing the coupled growth-transition process. In Sec.~\ref{section:results}, we will use our FPR model to obtain our main results about the multimodal and unimodal evolution modes for an array of islands and the role of asymmetric transitions in the evolution. In addition, we compare our FPR model with the existing Fokker-Planck model with discontinuous chemical potential. \cite{Vine2005} In Sec.~\ref{section:discussion}, we will discuss the meaning of our results for growth of quantum dot arrays, and compare our theoretical results with those in experiments and simulations. In Sec.~\ref{section:summary}, we summarize our findings.

\section{Mathematical Model} \label{section:model}
\subsection{Two-dimensional energy model for strained islands}\label{subsection:energy}
\begin{figure}[!ht]
\centering
\subfloat[]{\includegraphics[width=0.5\textwidth]{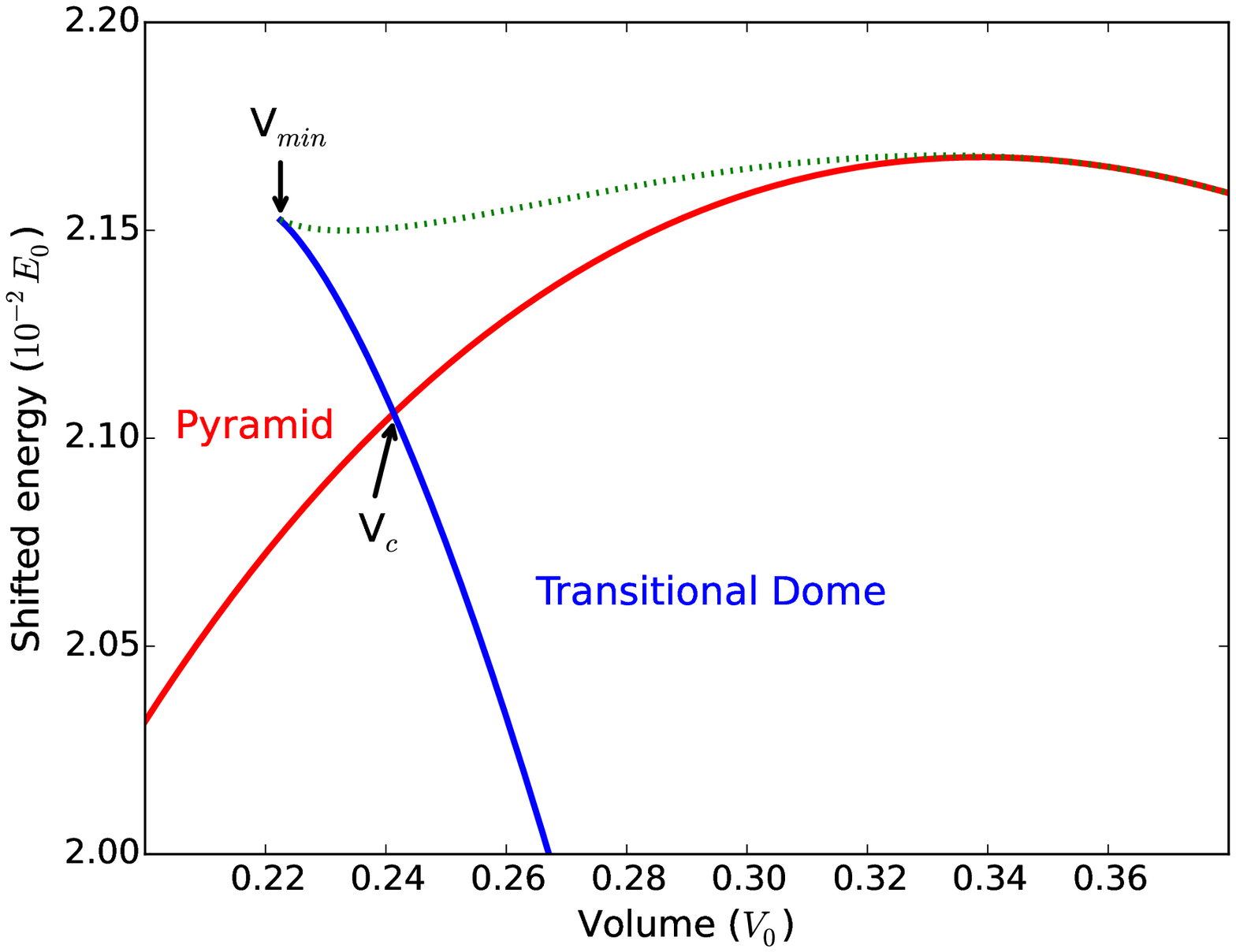}}
\subfloat[]{\includegraphics[width=0.5\textwidth]{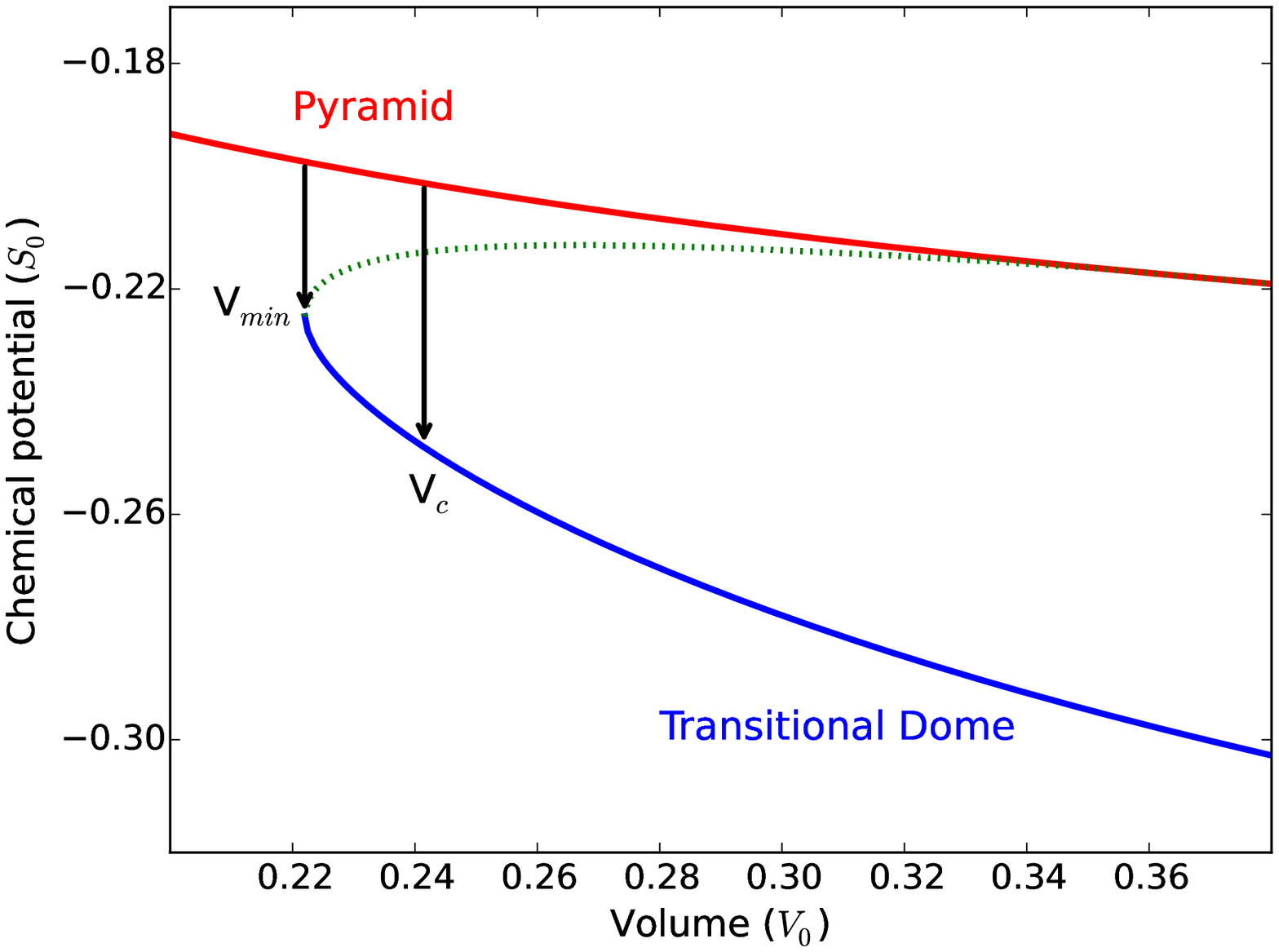}}
\caption{(Color online) (a) Energy curves and (b) chemical potentials of pyramid and transitional dome as a function of volume. The shifted energy $E+0.215S_0V$ is used for better resolution of the structure in (a). Dotted lines are unstable barrier solutions. $V_{min}$ is the minimal volume of existence of transitional dome and $V_c$ is the critical transition volume.} \label{fig:illustration}
\end{figure}

In this article, we consider a 2D system consisting of an array of Ge islands on a Si (001) substrate that undergoes epitaxial growth with the islands undergoing a sequence of shape transitions during growth. The array is assumed to be sparse so that the elastic interactions between islands can be neglected. We use the 2D energy model constructed in previous work \cite{Wei20160262} to describe the total energy including elastic energy and surface energy of a single 2D island. In this energy model a third-order approximation to the elastic energy is obtained using a perturbation method based on a thin-film and small-slope approximation
\begin{equation}
E_{el} = S_0V+S_0\int [-2hH(h_x)+4hH^2(h_x)-4hh_x^2]dx,
\end{equation}
where $V$ is the volume of island, $y=h(x)$ is the island shape function, $S_0$ is the elastic energy density of a planar film, and $H(f) = \frac{1}{\pi} \int_{-\infty}^{\infty} \frac{f(s)}{x-s} ds$ is the Hilbert transform. The surface energy is \begin{equation}
E_{surf} = \int\gamma(\theta)\sqrt{1+h_x^2}dx,
\end{equation}
where $\gamma(\theta)$ is the surface energy density as a function of the surface orientation $\theta$. Here we assume our 2D islands are fully-faceted with a small set of allowed facet orientations \{$\theta=m\theta_0|\ \theta_0=11.2^{\circ}$, $m=0, \pm 1, \pm 2, \pm 3$\}. This choice of facet orientations is a 2D analog for the $\{105\}(m=1)$, $\{113\}(m=2)$ and $\{15\ 3\ 23\}(m=3)$ facets on SiGe islands grown on Si(001). Thus we can classify different island types by a facet orientation sequence whose $i$-th element indicates the orientation index $m$ of the $i$-th facet of the island. For examples of symmetric island types, a pyramid (P) corresponds to (1,0,-1), a transitional dome (D$_1$) corresponds to (1,2,1,0,-1,-2,-1) and a multifaceted dome (D$_m$) corresponds to (1,2,3,2,1,0,-1,-2,-3,-2,-1). In addition to these symmetric island types, we also consider asymmetric island types such as half transitional dome H$_1$ (1,2,1,0,-1) and asymmetric transitional dome A$_1$ (1,2,3,2,1,0,-1,-2,-1), etc. Furthermore, for simplicity, we assume that all facets have the same surface energy density, i.e. $\gamma(\theta_i)=\gamma$. By minimizing the total energy $E_{tot}=E_{el}+E_{surf}$ with respect to the shape of a given island type subject to a prescribed volume constraint, we can find the equilibrium shape for each island type at any volume. Hence we get the energy as a function of the island volume for each island type, i.e. $E_I(V)$ with $I$ denoting the island type. \cite{Wei20160262} An illustrative example of these energy functions relating to the shape transition between pyramid and transitional dome (restricted to symmetric island shapes) is shown in Fig.~\ref{fig:illustration}. In Fig.~\ref{fig:illustration}(a), the energy curves of pyramid and transitional dome cross each other at the critical transition volume $V_c$, below which the pyramid is energetically favorable and above which the transitional dome is favorable. The two locally metastable solutions are connected by an unstable barrier solution indicated by the dotted line which corresponds to a symmetric shape intermediate between the pyramid and transitional dome. Thus the transition between pyramid and transitional dome is a first-order transition and there will be a jump in the chemical potential $\mu(V)=\partial E(V)/\partial V$ (see Fig.~\ref{fig:illustration}(b)) when a pyramid transforms into a transitional dome.\par
\begin{figure}[!ht]
\centering
\subfloat[]{\includegraphics[width=0.5\textwidth]{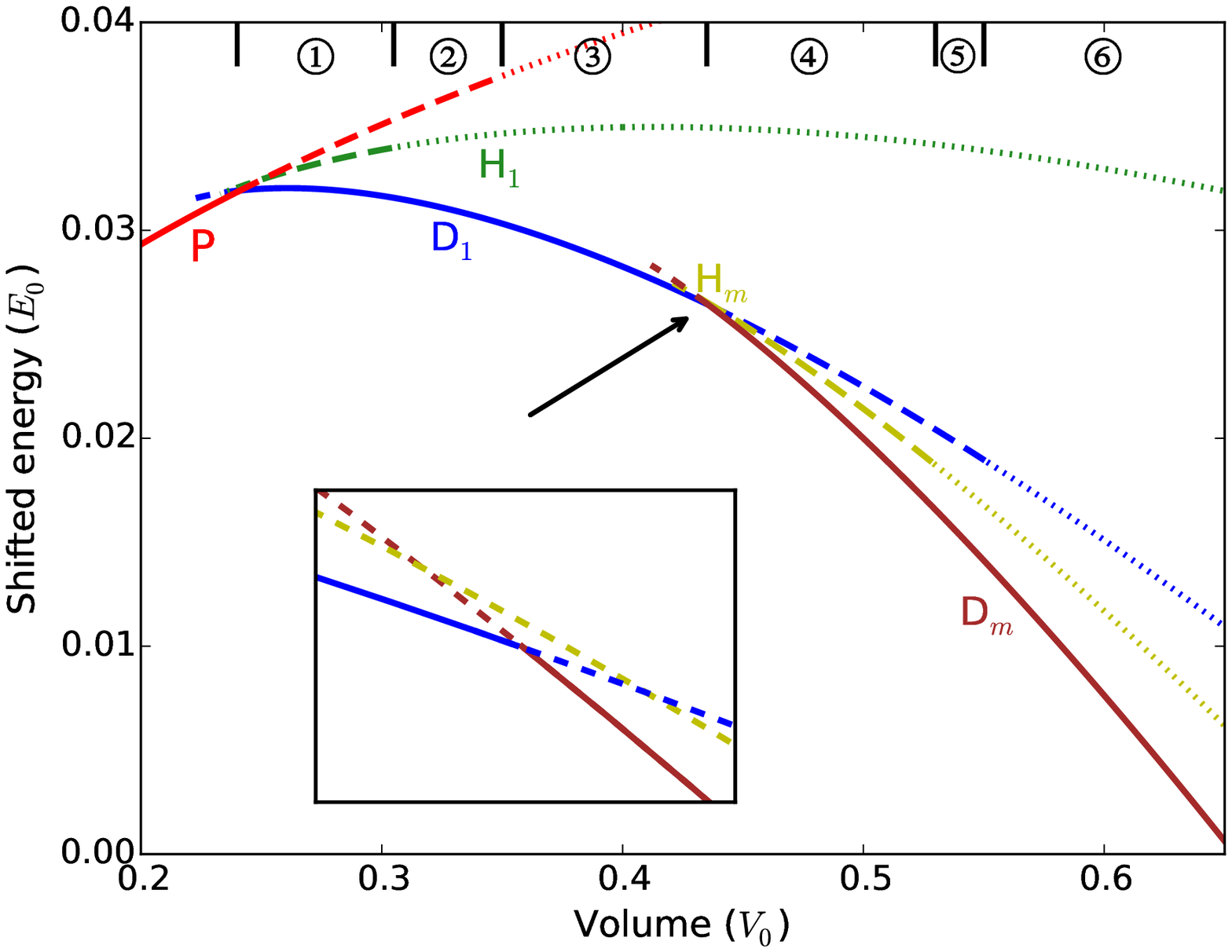}}
\subfloat[]{\includegraphics[width=0.5\textwidth]{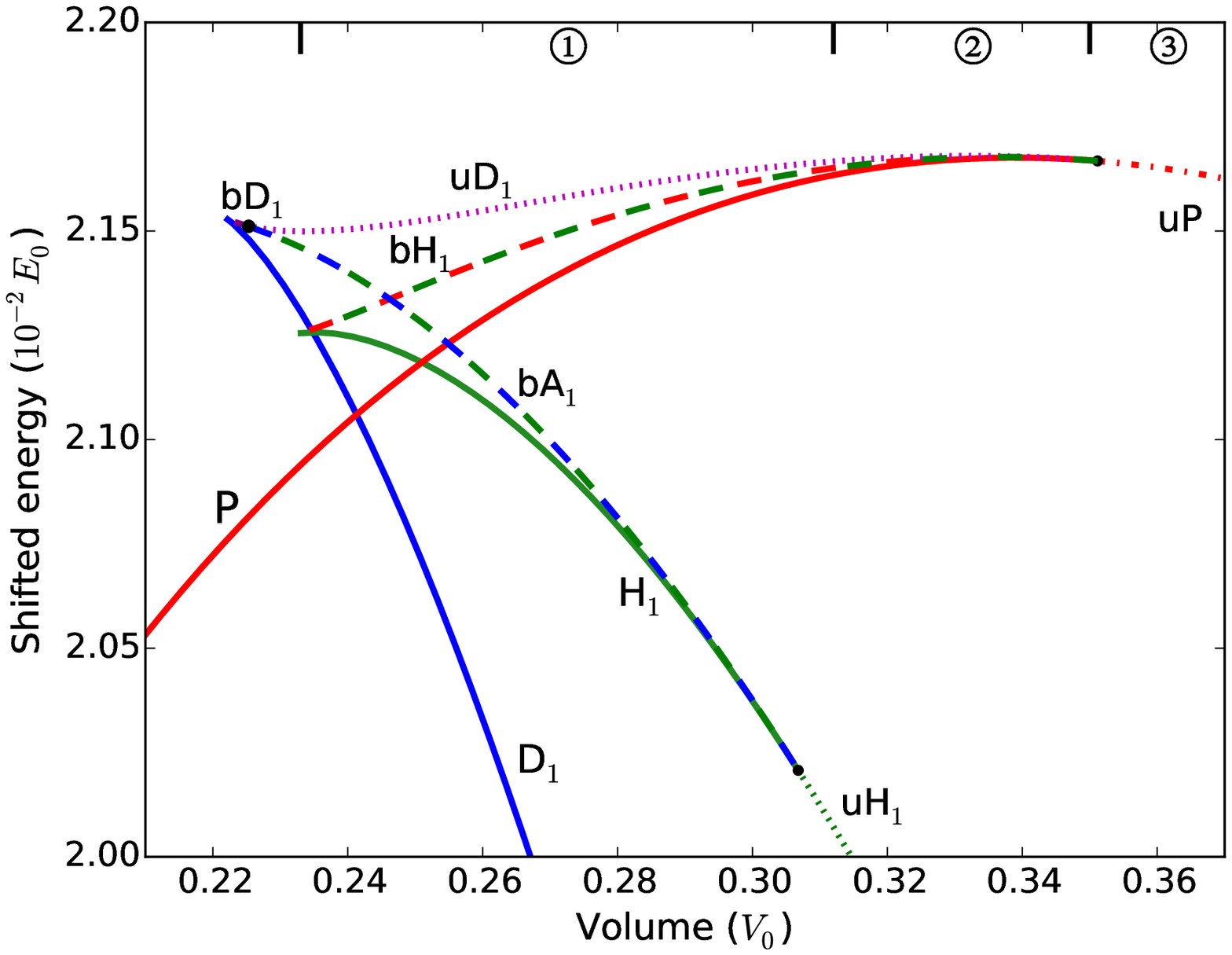}}
\caption{(Color online) (a) Energy curves of stable and metastable states, where the shifted energy is $E+0.26S_0V$. (b) Bifurcation diagram for energy of equilibrium states for P-D$_1$ transition, where the shifted energy is $E+0.215S_0V$. Solid curves: stable or metastable states; dashed curves: energy barrier (labeled by lowercase letter b); dotted curves: unstable local energy maxima (labeled by lowercase letter u). Notations: P~(pyramid), D$_1$~(transitional dome), H$_1$~(half transitional dome), A$_1$~(asymmetric transitional dome), D$_m$~(multifaceted dome), H$_m$~(half multifaceted dome). Numbers \textcircled{\footnotesize1}-\textcircled{\footnotesize6} at the top of figures label the volume ranges for different transition stages.} \label{fig:bifurcation}
\end{figure}

Based on this energy model, we determine the full results for the transitions of all island types, as summarized in Fig.~\ref{fig:bifurcation}. We have obtained two main conclusions: \cite{Wei20160262} (1) as shown in Fig.~\ref{fig:bifurcation}(a), pyramid, transitional dome and multifaceted dome are stable states (energetically favorable) for small, middle-sized and large islands respectively, so during growth an island will undergo a first-order transition with a jump in the chemical potential from a pyramid to a transitional dome, then to a multifaceted dome; (2) the lowest-barrier pathways of the two transitions are strongly dependent on the island volume and each transition can be divided into three volume ranges associated with three transition stages (see Sec.~\ref{subsection:reaction} for details). An important result is that the transition will involve asymmetric island types, such as half dome (H$_1$) (see Fig.~\ref{fig:bifurcation}(b)) and half multifaceted dome (H$_m$), as metastable states at the early transition stages. For example, considering an island at the critical transition volume ($V=0.2415$ in Fig.~\ref{fig:bifurcation}(b)), the P will not transform directly into D$_1$ through the symmetric barrier uD$_1$ (as shown in Fig.~\ref{fig:illustration}(a)), but instead it will firstly transform into a metastable asymmetric shape H$_1$ through a lower barrier bH$_1$ and then transform into D$_1$ through a second barrier bA$_1$. These barriers cause the transition to be governed by kinetics and control the dynamics of the transition. In general, the barriers decrease as the island volume increases, so the growing islands that have not transformed become more susceptible to transforming later. Thus, with multiple transition states and transition barriers that change with island size, the dynamics of a growing array of islands is complex. \par
To try to understand these complex dynamics, in this paper we will concentrate our discussion on the dynamics of the growth and transitions between the five stable or metastable island types P, H$_1$, D$_1$, H$_m$ and D$_m$. To study the coupled growth-transition process, we will assume that each type of island grows independently by the attachment or detachment processes, and that a given island may transform to a different island type with the same volume. This important assumption enables us to use a Fokker-Planck model to describe the growth process of each island type independently, and use reaction rates to describe the shape transition processes between different island types.

\subsection{The Fokker-Planck model for growth process} \label{subsection:growth}
Regarding the description of the growth process, we will use a kinetic approach based on the Fokker-Planck equation. The Fokker-Planck model arises as a continuum generalization of the reaction kinetic model of a discrete Becker-D\"{o}ring equation for describing nanocluster growth. \cite{risken1996fokker,Pirkkalainen2008twocomputational} A mean-field chemical potential for island growth was introduced in the Fokker-Planck model to describe the coarsening process of the quantum dots on a substrate with an external deposition of material in Refs.~\onlinecite{jesson2004metastability,Vine2005,Munt2007}. In their model, they assume that at each volume there is only one island type existing even when two island types are considered. \cite{Vine2005} In our paper, however, we will generalize this Fokker-Planck approach to describe the dynamics of multiple island types that may coexist at the same volume. \par
For a system consisting of an array of different types of islands, suppose $f_I(t,V)$ is the island size distribution function such that $f_I(t,V)dV$ specifies the number of islands of type $I$ per unit area with size between $V$ and $V+dV$ at time $t$ (where $I=P,H_1,D_1,H_m,D_m$ represents the island type corresponding to pyramid, half transitional dome, transitional dome, half multifaceted dome and multifaceted dome). If the system is at thermodynamic equilibrium, we have the detailed balance for the growth process of islands of type $I$
\begin{equation}
W_I(V\rightarrow V+dV)f_{I,eq}(V)dV=W_I(V+dV\rightarrow V)f_{I,eq}(V+dV)dV,
\end{equation}
where $W_I(V\rightarrow V+dV)$ and $W_I(V+dV\rightarrow V)$ are the corresponding growth/dissolution rates limited by the attachment or detachment processes respectively. The equilibrium distribution of islands of type $I$, $f_{I,eq}(V)$, is calculated using a statistical physics model which takes different island types in to account \cite{Rudd2003,Rudd2007}
\begin{equation}
f_{I,eq}(V)=A\exp\{\frac{\overline{\mu}V-E_I(V)}{k_BT}\},
\end{equation}
where $A$ is a normalization constant for the probability distribution, $k_B$ is the Boltzmann constant, $T$ is temperature measured in $K$, $E_I(V)$ is the total energy of an island of type $I$ at size $V$ obtained using the energy model in Sec.~\ref{subsection:energy}, and $\overline{\mu}$ is the mean-field chemical potential of the system. In particular, $\overline{\mu}$ serves as the chemical potential of the adatoms in the wetting layer reservoir and can be calculated by the conservation law for the total volume of the system (see Sec.~\ref{subsection:coupling}). In the non-equilibrium kinetic growth process, suppose $J_I(t,V)$ is the net flux of type-$I$ islands between $V$ and $V+dV$ and we have
\begin{eqnarray}
J_I(t,V)&=&W_I(V\rightarrow V+dV)f_I(t,V)dV-W_I(V+dV\rightarrow V)f_I(t,V+dV)dV \nonumber\\
&=&W_I(V\rightarrow V+dV)f_{I,eq}(V)dV[\frac{f_I(t,V)}{f_{I,eq}(V)}-\frac{f_I(t,V+dV)}{f_{I,eq}(V+dV)}] \nonumber\\
&=&-W_I(V\rightarrow V+dV) f_{I,eq}(V)(dV)^2\frac{\partial}{\partial V}[\frac{f_I(t,V)}{f_{I,eq}(V)}] \nonumber\\
&=&w_I(V)[\frac{\overline{\mu}(t)-\mu_I(V)}{k_BT}f_I(t,V)dV-\frac{\partial f_I(t,V)}{\partial V}dV]
\end{eqnarray}
where $w_I(V)=W_I(V\rightarrow V+dV)\cdot dV$ is the volume growth rate of a type-$I$ island at size $V$, and we use Eq.~(3) in the second line and Eq.~(4) in the last line. Note that although our $J_I(V)$ and $w_I(V)$ are defined with respect to a small interval of island volume [$V,V+dV$] rather than the number of atoms as in Ref.~\onlinecite{Munt2007}, the resulting system in Ref.~\onlinecite{Munt2007} can be recovered from our derivation. The flux $J_I(t,V)$ consists of two terms: the first term is the drift contribution and the second term is the diffusion contribution. Now considering the rate of change of the number of type-$I$ islands at size between $V_1$ and $V_2$ due to the growth process, we have
\begin{equation}
\frac{\partial}{\partial t}\int_{V_1}^{V_2} f_I(t,V) dV = J_I(V_1)-J_I(V_2) = -\int_{V_1}^{V_2} \frac{\partial J_I(t,V)}{\partial V} dV
\end{equation}
for any $V_1$ and $V_2$. Then we obtain the Fokker-Planck equation governing the growth process of islands of each type
\begin{equation}\label{eq:6}
\frac{\partial f_I(t,V)}{\partial t}=-\frac{\partial J_I(t,V)}{\partial V}.
\end{equation}\par

\subsection{Reaction terms for shape transition process} \label{subsection:reaction}
\begin{figure}[!ht]
\centering
\subfloat[]{\includegraphics[width=0.5\textwidth]{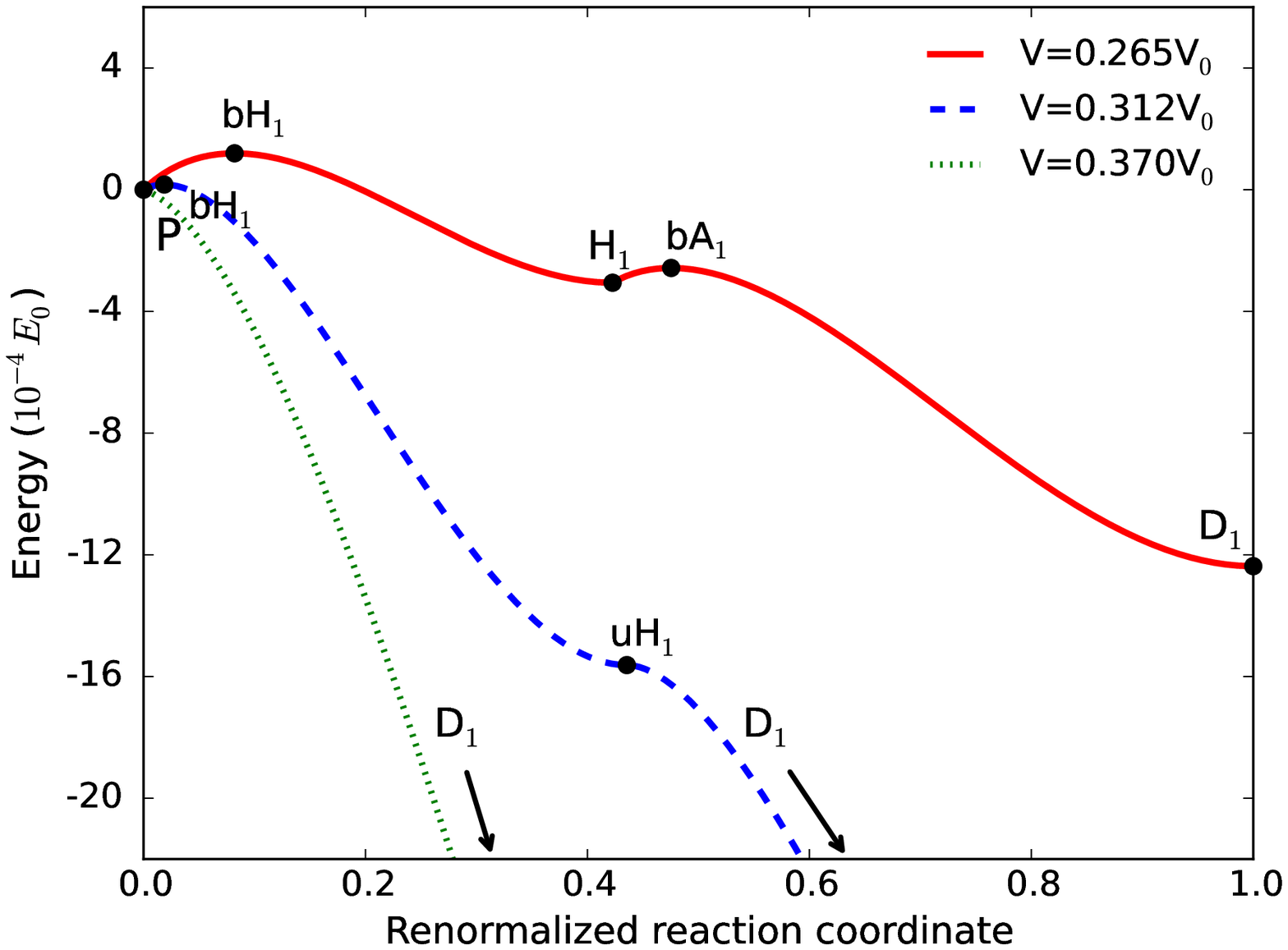}}
\subfloat[]{\includegraphics[width=0.5\textwidth]{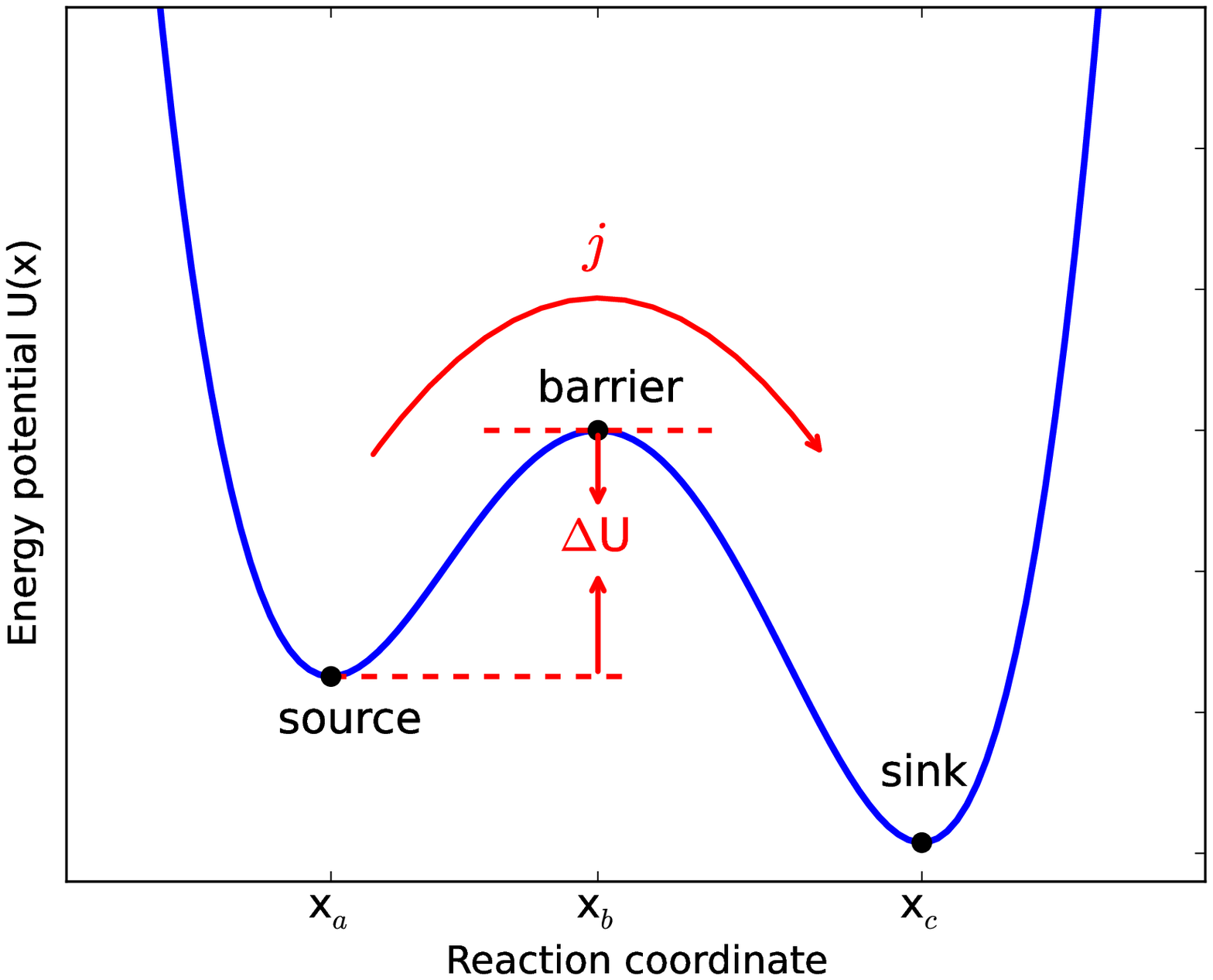}}
\caption{(Color online) (a) Energy of representative transition pathways for three stages of transition from pyramid to transitional dome. Notations are given in Fig.~\ref{fig:bifurcation}. (b) Energy potential $U(x)$ with two local minima $x_a$ and $x_c$, where a steady flux $j$ passes from $x_a$ to $x_c$ over a barrier $x_b$.} \label{fig:pathway}
\end{figure}

For the shape transition process, we focus on the shape transitions between islands of different types with the same volume. Based on the energy model in Sec.~\ref{subsection:energy}, we have constructed the lowest-barrier shape transition pathways as a function of island size and divided the transition process into several stages in Ref.~\onlinecite{Wei20160262}. There are three different transition stages, depending on the volume ranges indicated by numbers \textcircled{\small1}-\textcircled{\small6} at the top of Fig.~\ref{fig:bifurcation}(a), for both transitions P-D$_1$ (\textcircled{\small1}-\textcircled{\small3}) and D$_1$-D$_m$ (\textcircled{\small4}-\textcircled{\small6}). The energetics of representative transition pathways for the three stages of the P-D$_1$ transition are shown in Fig.~\ref{fig:pathway}(a). At the first stage \textcircled{\small1} ($V=0.265V_0$), the island sequentially nucleates two steep transitional facets (slope index of $\pm 2$) by climbing two barriers (bH$_1$ and bA$_1$) and goes through an one-sided metastable state H$_1$ along the transition path P-bH$_1$-H$_1$-bA$_1$-D$_1$; at the second stage \textcircled{\small2} ($V=0.312V_0$), the transition path P-bH$_1$-D$_1$ will not go through a metastable state and there is only one barrier bH$_1$; at the third stage \textcircled{\small3} ($V=0.37V_0$), the pyramid is unstable and the transition P-D$_1$ is symmetric and downhill in energy. The transition stages \textcircled{\small4}-\textcircled{\small6} are just the analog of \textcircled{\small1}-\textcircled{\small3} for the transition D$_1$-D$_m$. We can represent these transitions using reaction equations, for example, the transition path P-bH$_1$-H$_1$-bA$_1$-D$_1$ in \textcircled{\small1} has the reaction equation
\begin{eqnarray}
P \mathrel{\mathop{\rightleftarrows}^{\mathrm{k_1}}_{\mathrm{k_2}}} H_1 \mathrel{\mathop{\rightleftarrows}^{\mathrm{k_3}}_{\mathrm{k_4}}} D_1,
\end{eqnarray}
where $k_i$ for $i=1,2,3,4$ are the corresponding reaction rates. \par
To describe the transition process and calculate the reaction rates, we will use a classical model of a Smoluchowski equation \cite{Gardiner,Reactionrate1990} which describes a particle moving in a one-dimensional potential $U(x)$, where $x$ denotes the reaction coordinate. Supposing $\rho(x,t)$ is the probability density of the particle along the reaction coordinate, we have the Smoluchowski equation
\begin{equation}
\frac{\partial{\rho(x,t)}}{\partial t}=D\frac{\partial}{\partial x}\bigg(\frac{ U'(x)}{k_B T}\rho(x,t)\bigg)+D\frac{\partial^2\rho(x,t)}{\partial x^2},
\end{equation}
which is actually derived from a stochastic differential equation
\begin{equation}
dx=v_D(x)dt+\sqrt{2D}dW(t)
\end{equation}
where $v_D(x)=-\frac{DU'(x)}{k_B T}$ is the drift caused by the energy potential and the latter term is the random fluctuation. Now we apply this model to the shape transition process by letting $U(x)$ be the energy of states along the transition path from one island type to the other and setting the coordinate $x$ as the length of the transitional facets being added or removed in the shape transition. We calculate the reaction rate between two local minima of $U(x)$ at $x_a$ and $x_c$ by considering a steady probability current $j(x,t)=j$ passing from the source $x_a$ to the sink $x_c$ over the barrier $x_b$, as shown in Fig.~\ref{fig:pathway}(b). The stationary probability density $\rho_0(x)$ corresponding to the steady current $j$ satisfies
\begin{equation}
j=-D\frac{U'(x)}{k_B T}\rho_0(x)-D\frac{\partial\rho_0(x)}{\partial x},
\end{equation}
which can be easily solved with
\begin{equation}
\rho_0(x)=\frac{j}{D}e^{-\frac{U(x)}{k_B T}}\int_{x}^{x_c}e^{\frac{U(y)}{k_B T}}dy,
\end{equation}
which satisfies the absorbing boundary condition at the sink $\rho_0(x_c)=0$. Then the reaction rate $k$ is given by
\begin{equation}
k=\frac{j}{\rho_0(x_a)}=D e^{\frac{U(x_a)}{k_B T}}\bigg(\int_{x_a}^{x_c}e^{\frac{U(y)}{k_B T}}dy\bigg)^{-1}.
\end{equation}
Here we regard the reaction rate as the steady current divided by the stationary probability at one single point $x_a$ instead of the whole probability in the well of the local minimum as in the classical reaction rate theory \cite{Reactionrate1990} because in our model we only focus on the population of stable and metastable island states rather than the population including other intermediate states. Note that when the barrier is large enough, i.e. $\Delta U(x_a)=U(x_b)-U(x_a)\gg k_BT$, the formula in Eq.~(13) can approximated using Laplace's method to obtain
\begin{equation}
k\approx D\sqrt{\frac{-U''(x_b)}{2\pi k_BT}}e^{-\Delta U(x_a) /k_B T}.
\end{equation}
Moreover, the reaction rate we get in this way is self-consistent with the detailed balance for the transition process. Considering the transition between P and H$_1$ in Eq.~(8) for instance, since we have constructed the transition path at each volume, we can calculate the reaction rates for the forward transition and backward transition with the volume-dependent potential $U_V(x)$. It is easy to check that the reaction rates calculated by Eq.~(14)
\begin{eqnarray}
k_1(V)=D\sqrt{\frac{-U_V''(bH_1)}{2\pi k_BT}}e^{[E_P(V)-E_{bH_1}(V)]/k_B T}, \\
k_2(V)=D\sqrt{\frac{-U_V''(bH_1)}{2\pi k_BT}}e^{[E_H(V)-E_{bH_1}(V)]/k_B T},
\end{eqnarray}
and the equilibrium distribution given in Eq.~(4) satisfy the detailed balance equation at equilibrium
\begin{equation}
k_1(V)\cdot f_{P,eq}(V)=k_2(V)\cdot f_{H_1,eq}(V).
\end{equation}\par
It is worth mentioning that a more general Fokker-Planck (or Smoluchowski) model in a higher-dimensional space of arbitrary lengths for all facets would possibly give a full description of the growth and shape transition process involving arbitrary island shapes including all intermediate states. Rather than using this possible full description which involves complex high-dimensional calculations of the energy surface, we use a reduced reaction model with reaction rates containing the information of the reaction potential, which is essentially a projection of an island energy potential in a high-dimensional space of all facets onto a reduced space of the island volume and transitional facet length, to describe the transitions between the stable and metastable island states.

\subsection{The coupled Fokker-Planck reaction model} \label{subsection:coupling}
We can now couple the descriptions of the shape transition process and the growth process of an array of islands by introducing the reaction terms into the Fokker-Planck equations Eq.~(7), and have the Fokker-Planck reaction model
\begin{eqnarray}
\frac{\partial \textbf{f}}{\partial t}&=&-\frac{\partial \textbf{J}}{\partial V}+ \textbf{K}\cdot \textbf{f},
\end{eqnarray}
where $\textbf{f}=(f_P,f_{H_1},f_{D_1},f_{H_m},f_{D_m})$ is the size distribution vector for all types of islands, $\textbf{J}=(J_P,J_{H_1},J_{D_1},J_{H_m},J_{D_m})$ is the corresponding flux vector given in Eq.~(5) and $\textbf{K}(V)$ is the matrix of reaction rates obtained in the previous section for all transitions between different types of islands. If the system is isolated from the external environment, both total number and total volume of islands are conserved. When an external deposition flux of material is present, the total number of islands remains constant while the total volume of the system changes due to the deposition flux
\begin{eqnarray}\nonumber
\Phi&=& \sum_I\int_{0}^{\infty} \frac{\partial f_I(t,V)}{\partial t} V dV \\
&=& -\sum_I\int_{0}^{\infty} \frac{\partial J_I(t,V)}{\partial V} V dV + \sum_I \int_{0}^{\infty} \sum_J K_{I,J}(V)f_J(V) V dV\\
&=& -\sum_I[VJ_I(t,V)]_{0}^{\infty}+\sum_I\int_{0}^{\infty} J_I(t,V)dV,
\end{eqnarray}
where the summation index $I$ ranges over the different island types, $\Phi$ is the deposition flux per unit area of the system surface per unit time, and the boundary term in last line will vanish. Note the second term in Eq.~(19) vanishes because $\sum_I K_{I,J}=0$ for each $J$, and we have used integration by parts in Eq.~(20). We can use Eq.~(20) together with Eq.~(5) to calculate a self-consistent mean-field chemical potential $\overline{\mu}(t)$ to maintain the balance of the total volume of the system. In particular, when there is no external deposition, i.e. $\Phi=0$, $\overline{\mu}$ is the average of the chemical potential $\mu_I(V)$ over all the islands. Then Eq.~(20) together with Eq.~(18) and Eq.~(5) close the system. \par

There are two main differences between this coupled Fokker-Planck reaction model and the model in Refs.~\onlinecite{jesson2004metastability,Vine2005,Munt2007}. First, Refs.~\onlinecite{jesson2004metastability,Vine2005,Munt2007} use a model formula for the energy of QDs as a function of island size that is assumed to possess a minimum in formation energy per volume. This model formula can describe different island types by choosing different parameter coefficients but it cannot capture the change of the shape of same island type at varying size. In contrast, we use an island energy calculated numerically based on the exact shape of the island for each island type. This energy does not have a minimum in formation energy per volume for all island types within the volume range of interest. Second, our model includes reaction terms to describe the asymmetric shape transitions between different island types. It is worth noting that in Ref.~\onlinecite{Vine2005} the shape transition between pyramid and dome was also considered by using a discontinuous chemical potential with an abrupt decrease from pyramid to dome at the critical transition volume $V_c$ as shown in Fig.~\ref{fig:illustration}(b). In this way, they assume that the transition from pyramid to dome happens instantaneously and completes at the critical transition volume regardless of the realistic process of transition. Our model treats the transition as an energy-driven, kinetically-limited process which depends on the transition path, energy barriers and transition rates.\par

\section{Numerical Results} \label{section:results}
Supposing that an array of small pyramid islands with a Gaussian distribution with respect to volume form initially on the substrate, we will simulate the evolution of this array of islands during the growth-transition process by numerically solving the above Fokker-Planck reaction model. By considering different values for the growth parameters in the system, we are able to investigate the detailed dependence of the evolution behavior on the growth environment. Furthermore, we will compare the evolutions of the island shape distribution described by our Fokker-Planck reaction model (FPR) with that of the Fokker-Planck model with a discontinuous chemical potential (DCP). \cite{Vine2005} Note both the FPR model and the DCP model contain discontinuities in chemical potential but at different volumes and for different reasons. The discontinuities in chemical potential of FPR model are located at the smallest volume of existence of each island type (see $V_{min}$ of D$_1$ in Fig.~\ref{fig:illustration}(a) for example) below which the island type is no longer an energetic minimum, while in the DCP model the discontinuities lie at the critical transition volume of the P-D$_1$ and D$_1$-D$_m$ transitions to force the transitions happen at such volumes. In addition, the reaction coefficients in the FPR model are piecewise continuous within the volume range of each transition stage. To deal with these discontinuities in our numerical solutions of Eq.~(18), we use a finite volume method with a flux limiter strategy combining the Lax-Wendroff scheme and Godunov scheme for transport-like equations. \cite{Numerical2013} Furthermore at each time step, we will update the current mean-field chemical potential $\overline{\mu}(t)$ using the balance equation of the total volume of system Eq.~(20). \par
Assuming that the island growth is limited by the attachment and detachment of adatoms, the volume growth rate $w_I(V)$ is scaled with the island-substrate contact perimeter, e.g. $w_I(V)\sim \sqrt[3]{V}$ for a 3D island. In our model, however, the islands with 2D shapes in a vertical plane are assumed to have thickness which extends freely in a third direction, so we take the thickness of the islands as the characteristic length in the simulations. Hence the volume growth rate for our islands is scaled with the number of the island edges ($=2$) contacting with the substrate and we can reasonably suppose the volume growth rate is uniform for all islands, i.e. $w_I(V)=w_0$. Our results for the simulations are shown in terms of characteristic scales for length $l_0=\gamma /S_0$, volume $V_0=l_0^3$, energy $E_0=V_0S_0$ and time $t_0=V_0/w_0$. Keeping the notations of all variables, we have the nondimensionalized system
\begin{eqnarray}
&&\frac{\partial \textbf{f}}{\partial t} = -\frac{\partial \textbf{J}}{\partial V}+ r\ (\textbf{K}\cdot \textbf{f}),\\
&&J_I(V) = C\cdot[\overline{\mu}(t)-\mu_I(V)]f_I(t,V)dV-\frac{\partial f_I(t,V)}{\partial V}dV, \\
&&\Phi = -\sum_I[VJ_I(t,V)]_{V_{min}}^{V_{max}}+\sum_I\int_{V_{min}}^{V_{max}} J_I(t,V)dV,
\end{eqnarray}
where $r=D/(w_0/V_0)$ represents the ratio of the typical (or proper) transition rate and growth rate, and $C=E_0/(k_BT)$ is the energy coefficient. Using the parameters for SiGe films (for Si: $a = 5.4310\times10^{-8}$ cm, $E = 13.0\times10^{11}$ erg/cm$^3$, $\nu=0.278$; for Ge: $a = 5.6575\times10^{-8}$ cm, $\gamma = 1927$ erg/cm$^2$) these characteristic scales are $l_0\sim 16$nm, $E_0\sim 5\times 10^{-9}$erg $\sim 3\times 10^{3}$eV.

\subsection{Multimodal and unimodal evolution modes of islands shape distribution} \label{subsection:regime}
\begin{figure}[!ht]
\centering
\subfloat[]{\includegraphics[width=0.5\textwidth]{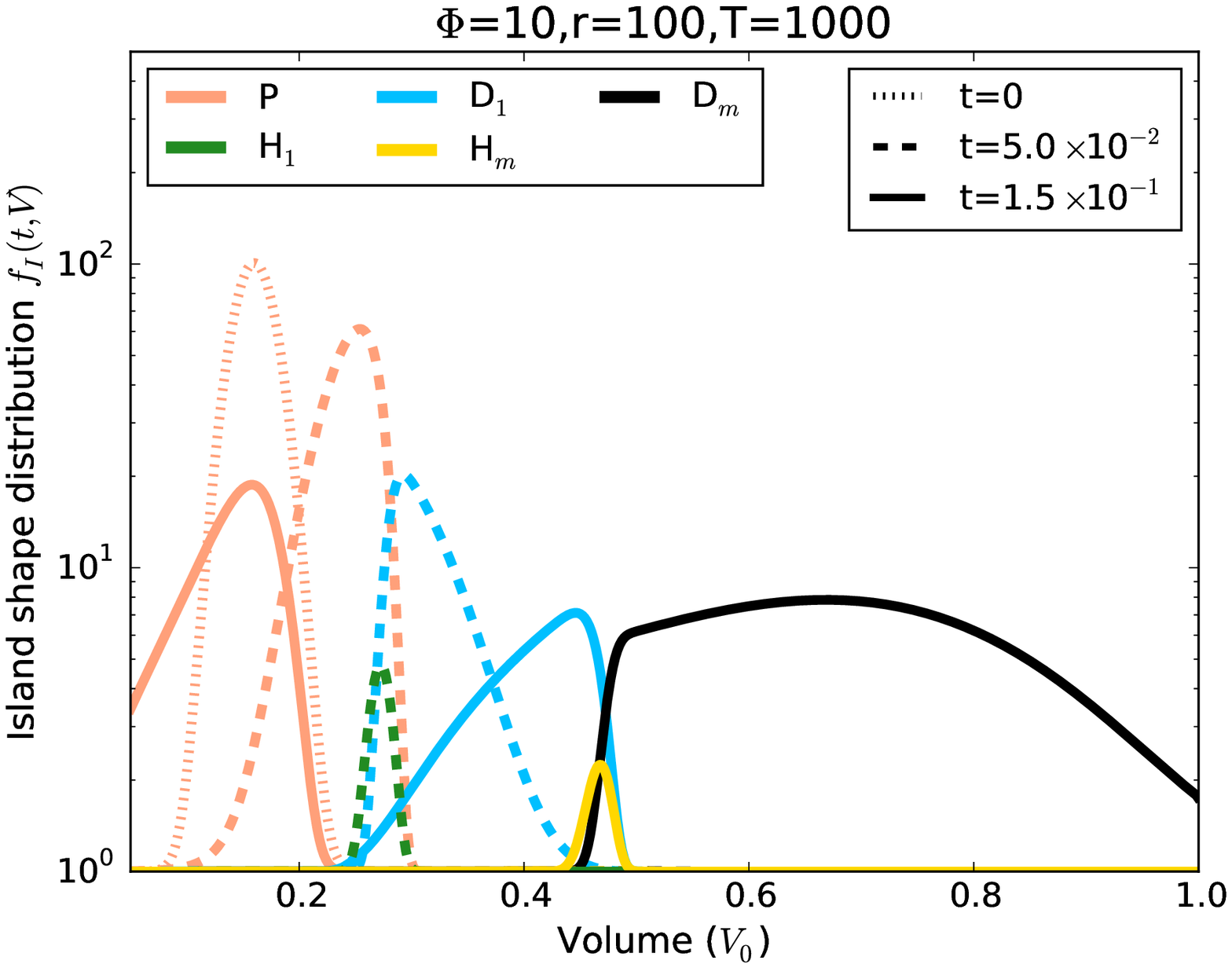}}
\subfloat[]{\includegraphics[width=0.5\textwidth]{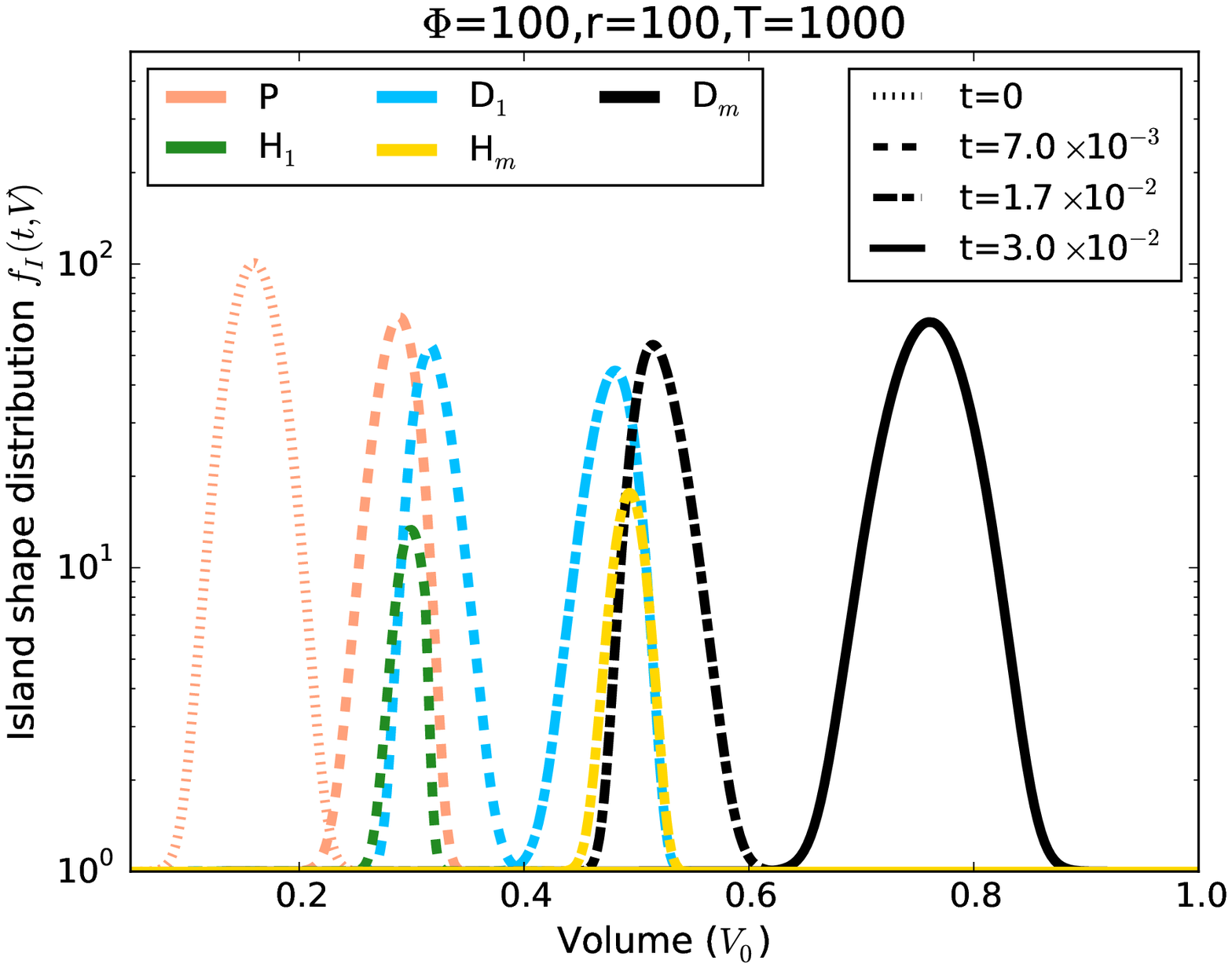}}\\
\subfloat[]{\includegraphics[width=0.5\textwidth]{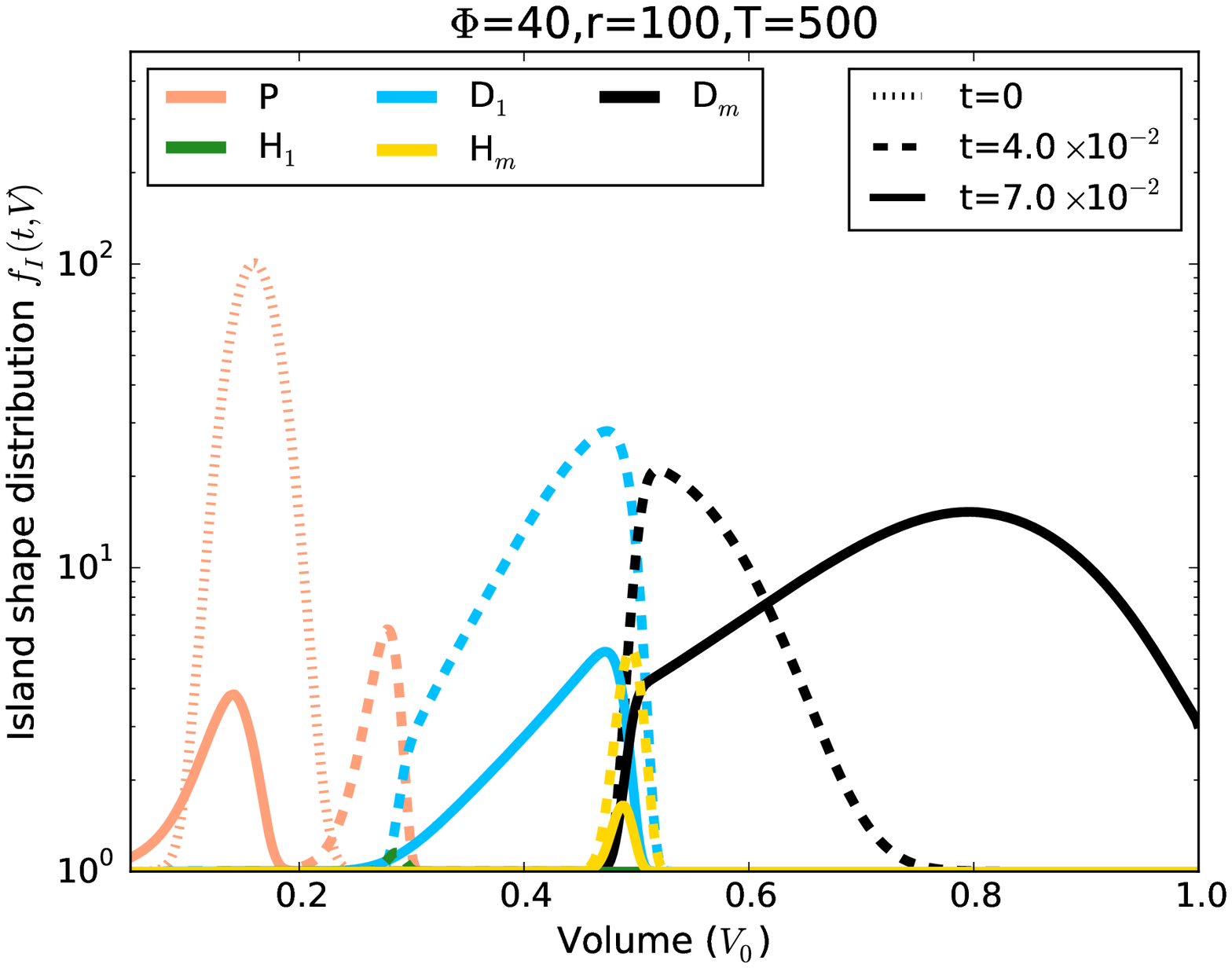}}
\subfloat[]{\includegraphics[width=0.5\textwidth]{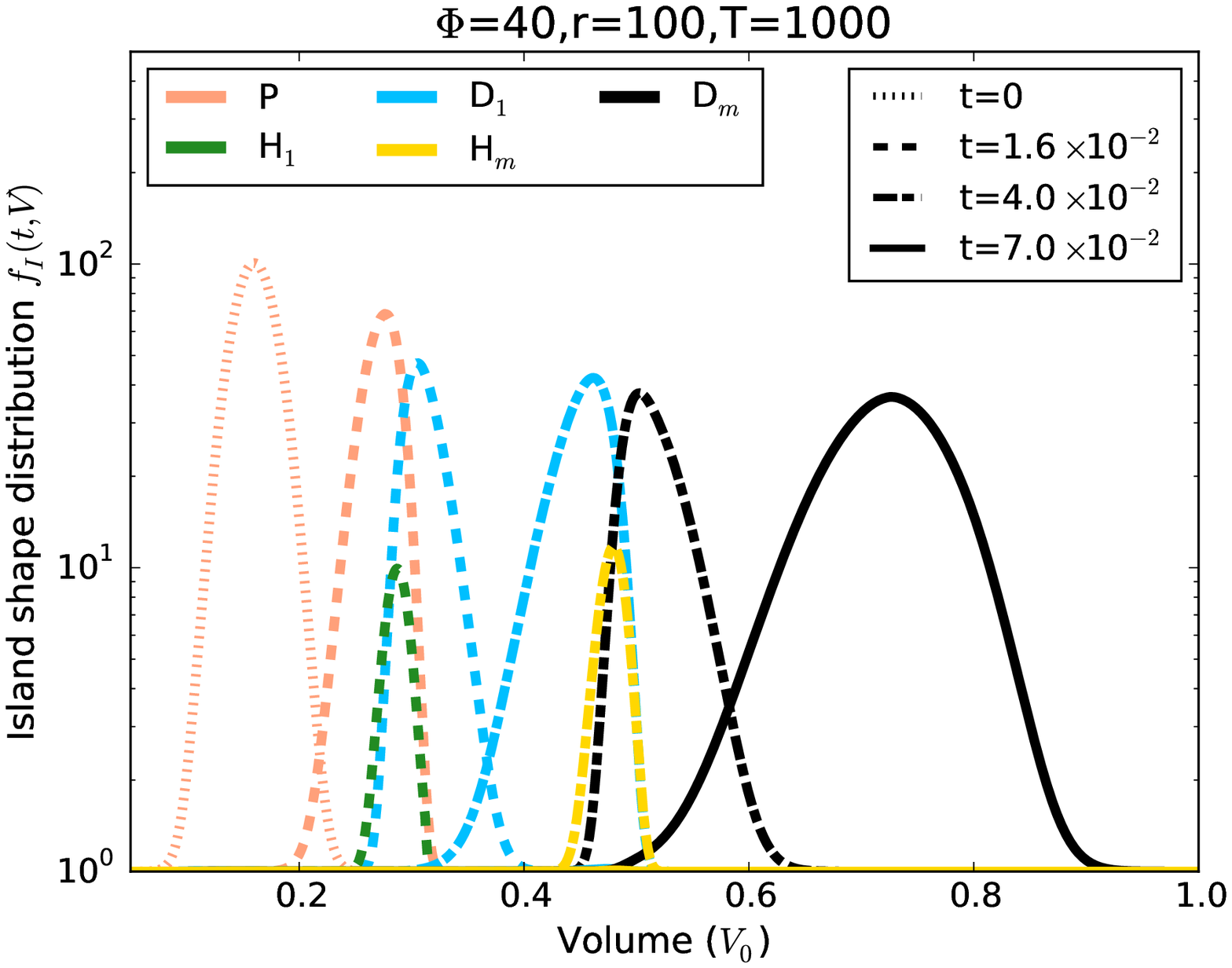}}
\caption{(Color online) Evolution of shape distributions in the multimodal mode (a), (c) and the unimodal mode (b), (d). Colored curves represent different island types as labeled. } \label{fig:regimes}
\end{figure}

Since we use a mean-field model to consider a dilute system of quantum dots, the magnitude of the island density is arbitrary. Letting $N_0/A_0$ be the characteristic density, we consider an array of pyramids of density $5N_0/A_0$ normally distributed with mean size $0.16V_0$ and standard deviation $0.02$. We will simulate the evolution of the island array under different growth conditions. We mainly focus on three growth parameters in our simulations: the external deposition flux rate $\Phi$ (measured in $w_0\cdot N_0/A_0$), the ratio of the shape transition rate and the island growth rate $r$, and the temperature $T$ (measured in $K$). By varying these three parameters, we find two typical evolution modes for the array of islands: multimodal and unimodal evolution modes. Both modes include the presence of asymmetric islands as part of the shape transition process, to be discussed in Sec.~\ref{subsection:asymmetry}. The phenomenon of the bimodal (multimodal) distribution of pyramids and domes has been observed in experiments \cite{Medeiros1998,Ross1998Coarsening,ross1999transition,Rastelli2005} and studied in theories. \cite{Ross1998Coarsening,daruka1999shape,Rudd2003,Jesson2004Suppression,Vine2005,lam2010kinetic} The multimodal (bimodal) distribution phenomenon is considered as a result of the first-order shape transition from pyramids to transitional domes after which the coarsening of islands occurs. \cite{daruka1999shape,Vine2005} This coarsening process is driven by the difference between the island chemical potential and the mean-field chemical potential $\overline{\mu}$ of the system, where pyramids possessing chemical potentials higher than $\overline{\mu}$ shrink while transitional and multifaceted domes possessing chemical potentials lower than $\overline{\mu}$ continue growing. However, we find that the island coarsening will be prohibited if $\overline{\mu}$ is kept higher than the chemical potentials of all islands during the evolution so that all islands will grow and transform into multifaceted domes resulting a unimodal distribution. As shown in Fig.~\ref{fig:regimes}(a-b) for a fixed ratio $r$ and temperature $T$, small deposition flux causes a multimodal distribution of pyramids P, transitional domes D$_1$ and multifaceted domes D$_m$, while large deposition flux surprisingly results in a unimodal distribution of multifaceted domes D$_m$. When a large flux of material is deposited into the system, the large number of new adatoms with high chemical potentials cause a reservoir with high mean-field chemical potential to balance the total volume of system, which drives the adatoms to attach to the existing islands including pyrmaids and domes. Moreover, as shown in Fig.~\ref{fig:regimes}(c-d), we find that for a fixed deposition flux rate, increasing the temperature of the system will also cause a transition of the evolution from the multimodal mode to the unimodal mode.\par
\begin{figure}[ht]
\centering
\subfloat[]{\includegraphics[width=0.5\textwidth]{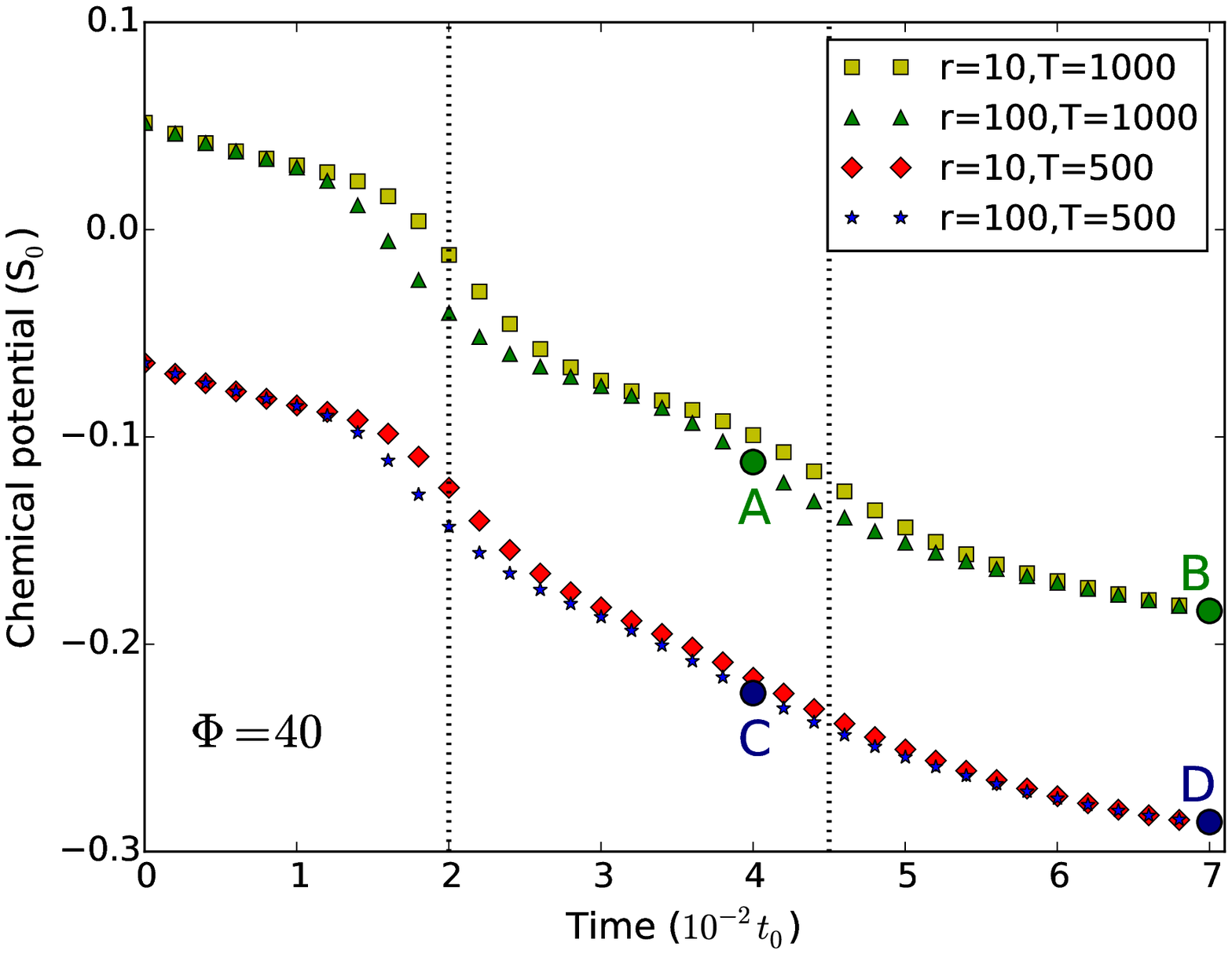}}
\subfloat[]{\includegraphics[width=0.5\textwidth]{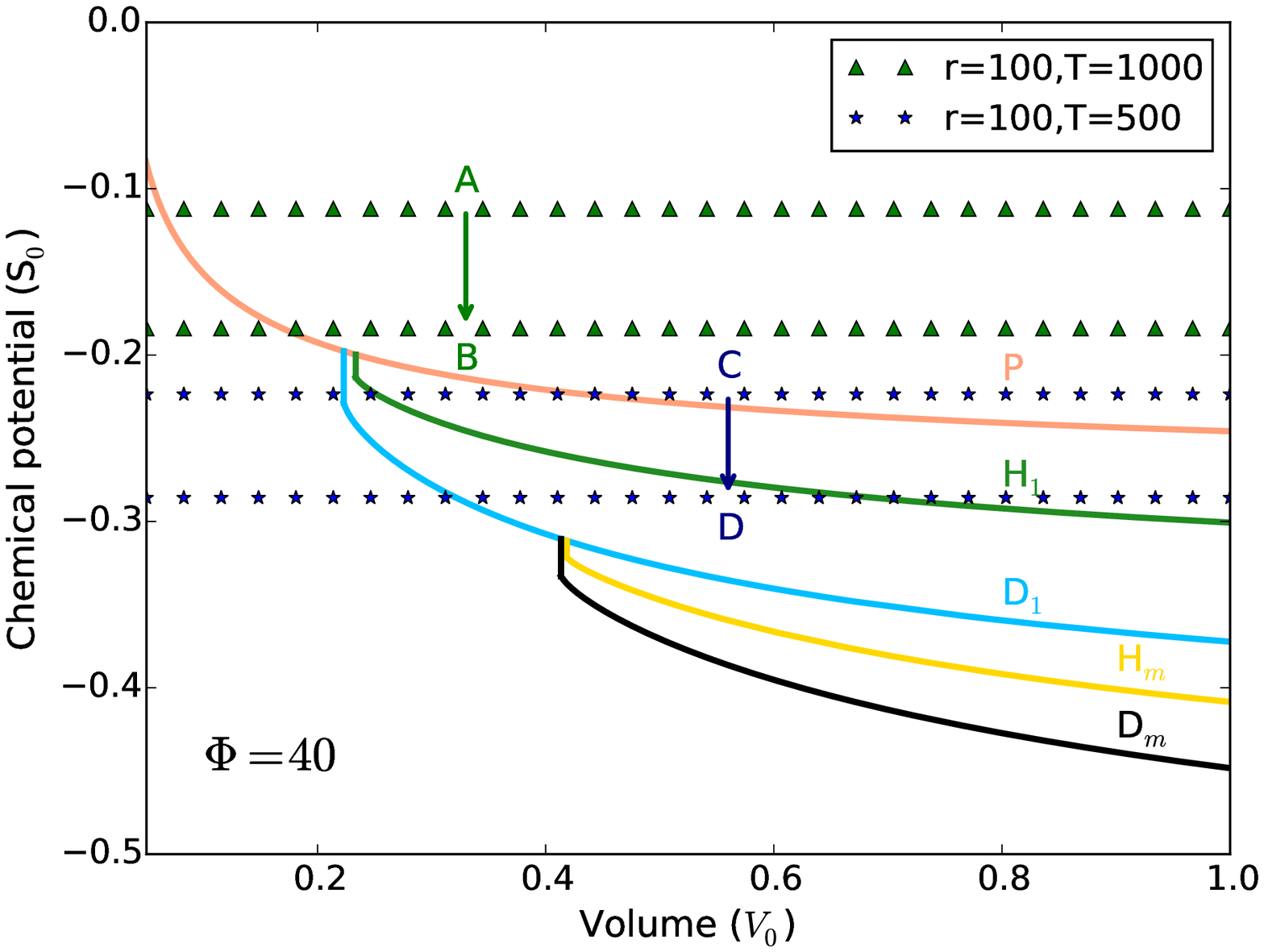}}
\caption{(Color online) (a) Evolution of the mean-field chemical potential $\overline{\mu}$ over time for $\Phi=40$. Vertical lines indicate the region of rapid decrease due to shape transitions. (b) Snapshots of mean-field chemical potential (horizontal symbol lines) and chemical potentials of different island types (colored solid curves). The symbol markers A, B, C and D represent the snapshots of $\overline{\mu}$ corresponding to the time points A, B, C and D during the evolutions in (a). Arrows indicate the decrease in $\overline{\mu}$ over time. Colors of solid curves represent different island types as labeled in (b).} \label{fig:evolution}
\end{figure}

To understand the detailed effects of $r$ and $T$ on the evolution mode, we trace the change of the mean-field chemical potential $\overline{\mu}$ of system during the evolution. Fig.~\ref{fig:evolution}(a) shows the change over time of two pairs of curves for $\overline{\mu}$, the upper pair corresponding to $T=1000K$ and the lower one corresponding to $T=500K$. During the evolution, $\overline{\mu}$ will decrease at both temperatures $T=500K$ and $T=1000K$ due to the island growth. However, $\overline{\mu}$ at $T=1000$ is always higher than the chemical potential of all the existing islands in the distribution hence resulting in a unimodal distribution, while $\overline{\mu}$ at $T=500$ will cross with the chemical potential of existing islands (mostly pyramids) hence resulting a multimodal distribution. To illustrate this phenomenon, we plot in Fig.~\ref{fig:evolution}(b) the snapshots of $\overline{\mu}$ at $t=4\times 10^{-2}$ (horizontal lines A, C) and $t=7\times 10^{-2}$ (horizontal lines B, D) together with the chemical potential curves of different island types. It shows that at high temperature $T=1000K$, although A and B intersect with the chemical potential of pyramid at small volumes, they will force all the existing islands grow (see Fig.~\ref{fig:regimes}(d)); whereas at lower temperature $T=500K$, C and D cross with the chemical potentials at a larger volume dividing the existing islands into two groups (see Fig.~\ref{fig:regimes}(c)), where the group of small islands (all pyramids, small half transitional domes and small transitional domes) shrink, while the group of larger islands continue growing. In fact, since our $\overline{\mu}$ is self-consistently calculated to make the balance equation for the total volume of system Eq.~(20) hold, for a fixed $\Phi$ a larger $\overline{\mu}$ is necessary when $T$ is higher, which can be directly observed from the form of the net flux $J_I(t,V)$ in Eq.~(5). Moreover, looking at the change of $\overline{\mu}$ in Fig.~\ref{fig:evolution}(a), there is an obvious rapid decrease of $\overline{\mu}$ around $t=0.02$ (vertical dotted line) at both $T=500$ and $T=1000$ and another rapid decrease around $t=0.045$ (vertical dotted line) at $T=1000$. The rapid decrease is because the islands are undergoing an abrupt jump in chemical potential due to the shape transitions P-D$_1$ around $t=0.02$ and D$_1$-D$_m$ around $t=0.045$. However, the D$_1$-D$_m$ transition does not cause an obvious rapid decrease in $\overline{\mu}$ around $t=0.045$ for $T=500$ because at this temperature the island evolution is in the multimodal mode where the increase of the chemical potential of shrinking pyramids obscures the rapid decrease caused by the shape transition. In particular, the decrease in $\overline{\mu}$ caused by the shape transition is faster when $r=100$ than it is when $r=10$, which reflects the effect of the transition/growth ratio $r$ on the evolution process: a larger $r$ can make the transition happen relatively faster but it will not affect the evolution mode of shape distribution. Thus we conclude that the evolution mode mainly depends on the deposition flux rate and the temperature of the system but not the transition rate $r$. \par
\begin{figure}[ht]
\centering
\includegraphics[width=0.5\textwidth]{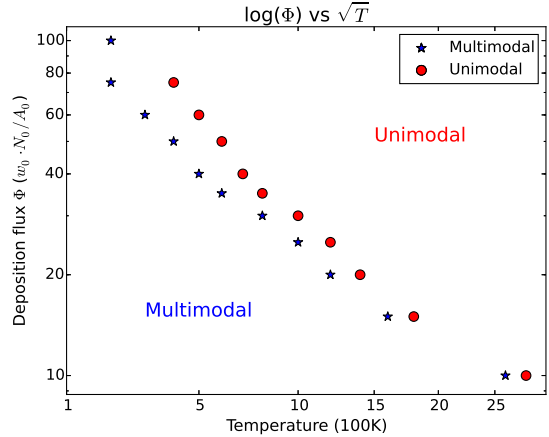}
\caption{(Color online) Transition boundary between the multimodal and the unimodal evolution modes in the parameter plane. The deposition flux rate axis is logarithmic scaled and the temperature axis is square root scaled. The transition/growth ratio is $r=100$. } \label{fig:parameter}
\end{figure}

From multiple simulations, we determine the parameter ranges separating the multimodal evolution mode and the unimodal evolution mode. Fig.~\ref{fig:parameter} shows the deposition flux~--~temperature parameter plane divided into two regions, where low flux, low temperature corresponds to the multimodal evolution mode and high flux, high temperature corresponds to the unimodal evolution mode. Fig.~\ref{fig:parameter} summarizes our main result of the evolution mode of the shape distribution based on our FPR model, that the evolution of the island shape distribution can change from a multimodal mode to a unimodal mode by either increasing the external deposition flux rate or increasing the temperature of the system.

\subsection{Comparison with Fokker-Planck model with discontinuous chemical potential (DCP)} \label{subsection:comparison}
\begin{figure}[ht]
\centering
\subfloat[]{\includegraphics[width=0.5\textwidth]{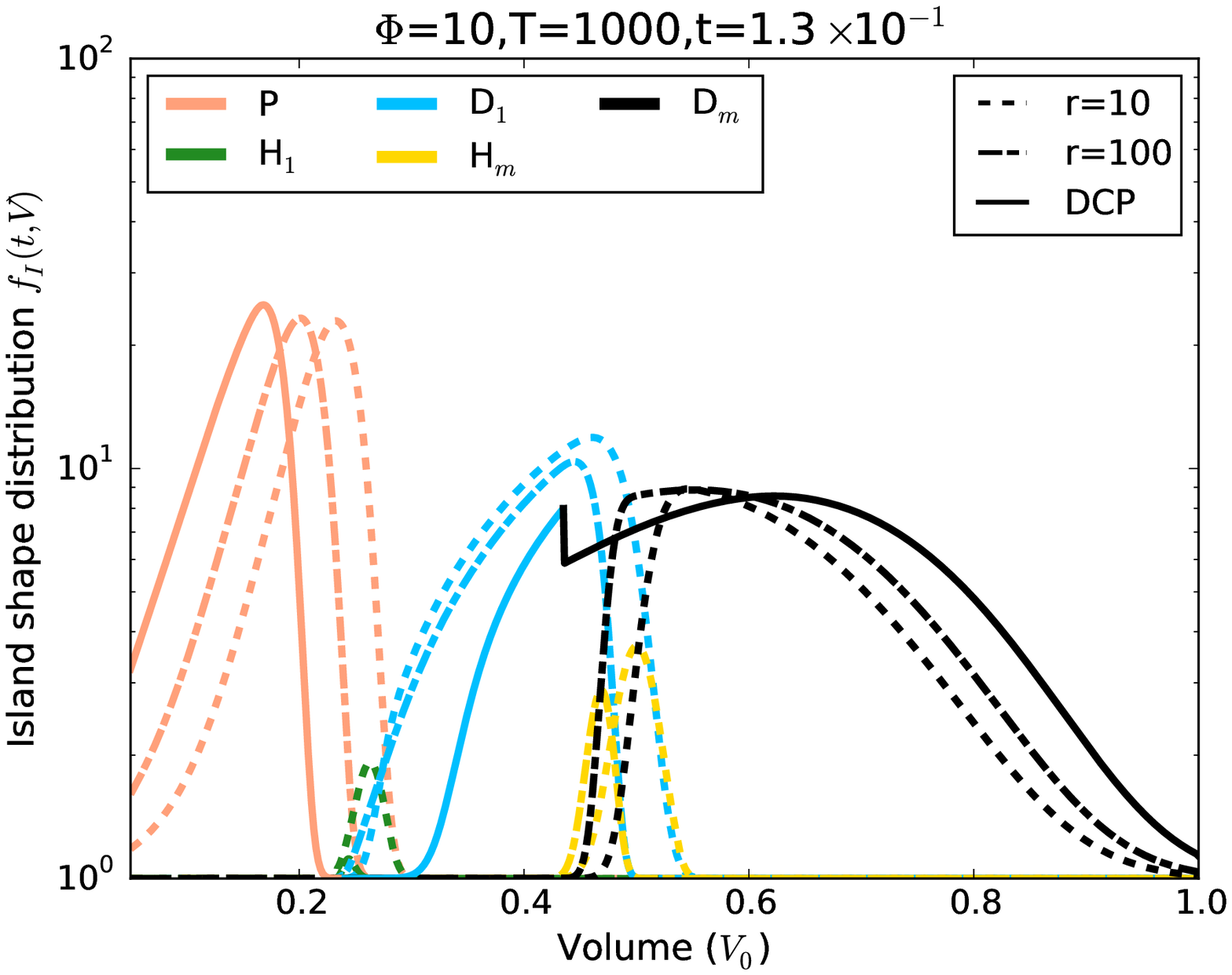}}
\subfloat[]{\includegraphics[width=0.5\textwidth]{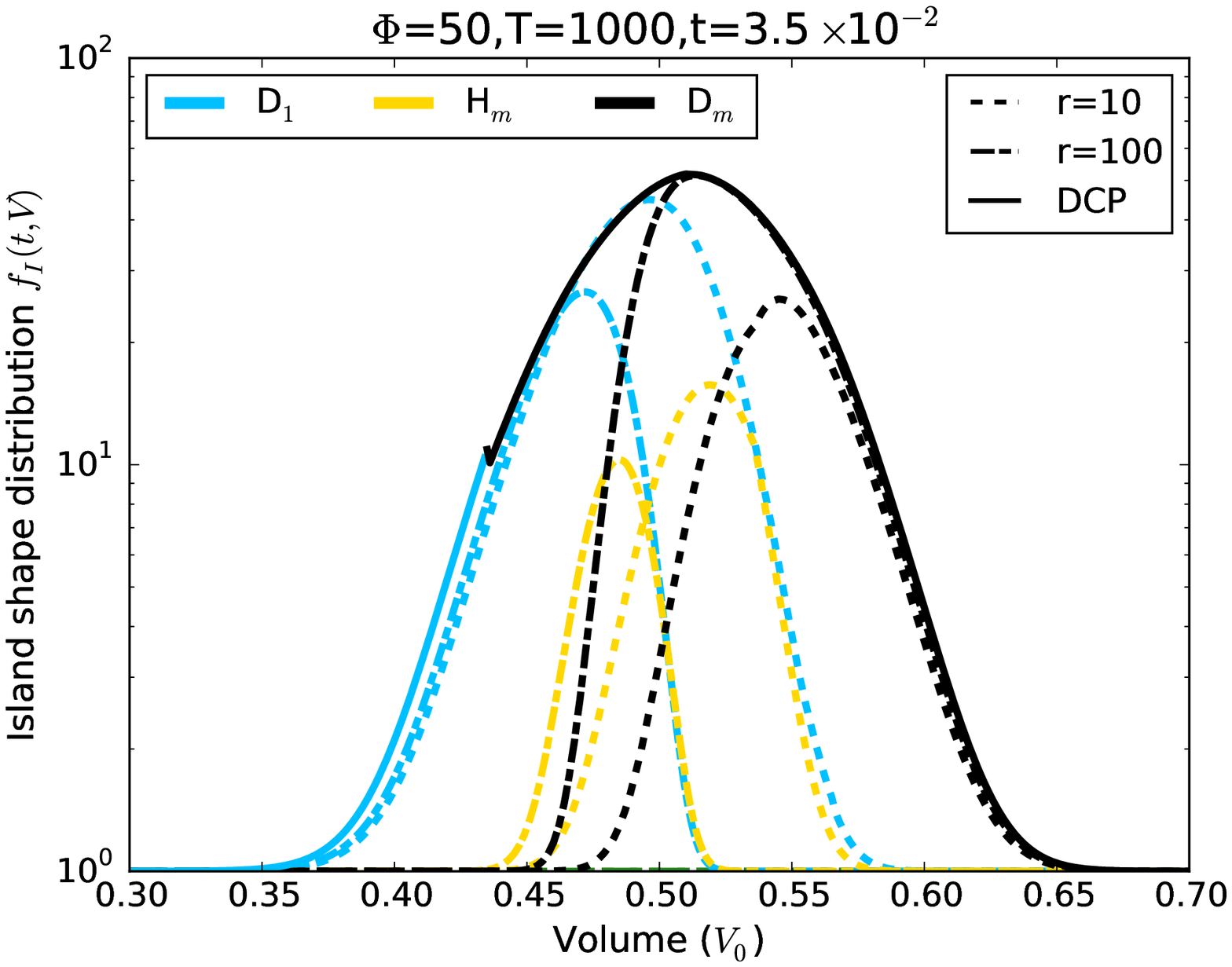}}
\caption{(Color online) Comparison of shape distribution results between the FPR model with $r=10,100$ and the DCP model. Colors represent different island types as labeled.} \label{fig:comparison}
\end{figure}

In our coupled Fokker-Planck reaction model, the reaction terms are introduced to describe the dynamics of the shape transition process where the transition/growth ratio $r$ controls the relative speed of the shape transitions. The DCP model \cite{Vine2005} discussed earlier in Sec.~\ref{subsection:coupling} uses a piecewise-defined chemical potential which enforces the pyramid to dome transition to occur instantaneously at a critical island volume. Thus the DCP model can be regarded a special case of our FPR model for $r\rightarrow\infty$ where the shape transitions between different island types are assumed to happen instantaneously. Since in Sec.~\ref{subsection:regime} we find that the ratio $r$ does not affect the island evolution mode, we expect that the DCP model will capture the similar parameter dependence of the evolution mode as shown in Fig.~\ref{fig:parameter}. We plot in Fig.~\ref{fig:comparison} the snapshots of the evolution of shape distributions in a multimodal mode ($\Phi=10$) and in a unimodal mode ($\Phi=50$) when $T=1000$ with different values of the ratio $r=10,100,\infty$. We observe that the DCP model does show the same evolution modes under different growth parameters as our FPR model predicts for the island shape distribution. And when $r$ is smaller, the shape transitions are slower so that the shrinking of pyramids (red curve) delays in Fig.~\ref{fig:comparison}(a) and there are more domes (blue curve) waiting to transform into multifaceted domes (brown curve) in Fig.~\ref{fig:comparison}(b). Although the coexistence of different island types in the multimodal evolution mode (Fig.~\ref{fig:comparison}(a)) is obviously seen for both DCP model and FPR model, the coexistence of different island types at a same size in the unimodal evolution mode (Fig.~\ref{fig:comparison}(b)) can only be captured by the FPR model. In other words, the FPR model can indicate the nonuniformity in island types and provide the detailed shape distribution profile even when the islands display an unimodal distribution in size. Moreover, the FPR model can show the evolution of the distributions of asymmetric islands during the shape transition process such as half transitional domes (green curve) and half multifaceted domes (yellow curve). More detailed results about asymmetric islands and asymmetric shape transitions will be given in next section.

\subsection{Asymmetric island and asymmetric transition effects} \label{subsection:asymmetry}
Although in our model we assume that the quantum dots are single-component material, alloyed materials are often deposited in the formation of quantum dots. A key feature of alloy quantum dots is that the surface composition can be nonuniform, \cite{Spencer2001Morphological,Tersoff2007Coarsening} and during growth or shape transitions the compositional nonuniformity becomes buried inside the island. \cite{alloiedprepyramid,Han2010composition,Shenoy2011} Thus to better utilize the self-assembly of quantum dots for the manufacture of nanostructured devices, not only the shape and size of the quantum dots are important but also the composition profile within the quantum dots are essential for the electronic properties of the nanostructured devices. In Sec.~\ref{subsection:energy}, we predict that the shape transitions between different island types during the growth can be asymmetric, which will result in asymmetric composition profiles within alloyed islands and subsequently affect the electronic properties of the nanostructured devices. Thus even for an array of islands in the unimodal evolution mode (uniform in size and shape), their electronic properties may vary significantly due to the nonuniform interior composition profile caused by different shape transition paths such as the asymmetric transition path and symmetric transition path. \par
In our FPR model, we introduce a reaction term with transition rates calculated based on the lowest-barrier transition path which enables us to take into account the effects of asymmetric shape transitions on the distribution evolution. In fact, we can trace the shape transition flux between different island types during the evolution to divide the transitional domes and multifaceted domes into two groups: ones from the symmetric shape transition and the others from the asymmetric shape transition. Considering an array of islands in the unimodal evolution mode with different deposition flux rates $\Phi$ and transition/growth ratios $r$, we calculate the corresponding final fraction of islands from asymmetric shape transition for transitional domes and multifaceted domes as shown in Table.~\ref{table}.
\begin{table}[tbp]
\centering
\caption{Fraction of islands passing through asymmetric shape transition for transitional domes (D$_1$) and multifaceted domes (D$_m$), where the temperature $T=1000K$.}
\label{table}
\begin{tabular}{|l||c|c||c|c||c|c||}
\hline
 deposition flux & \multicolumn{2}{c||}{$\Phi$ = 100} & \multicolumn{2}{c||}{$\Phi$ = 75} & \multicolumn{2}{c||}{$\Phi$ = 40} \\ \hline
\backslashbox{ratio}{type}  & D$_1$         & D$_m$          & D$_1$          & D$_m$          & D$_1$          & D$_m$        \\
\hline
$r = 100$       &    100\%      &     100\%      &      100\%     &      100\%     &     100\%      &     100\%      \\ \hline
$r =10$         &    62.6\%     &     84.2\%     &      74.2\%    &      91.9\%    &     94.5\%     &     99.4\%    \\ \hline
$r = 1$         &    8.9\%      &     14.6\%     &      11.9\%    &      20.2\%    &     21.9\%     &     37.9\%    \\ \hline
\end{tabular}
\end{table}
In general, for a fixed deposition flux rate, as the ratio $r$ decreases, the fraction of asymmetric transition islands for both transitional domes and multifaceted domes will decrease and symmetric transition of islands become dominant (see along columns in Table.~\ref{table}). Since $r$ represents the ratio of typical transition rate and growth rate, it describes the competition between the kinetic limitations of shape transition and island growth. When $r$ is large, the shape transition is less kinetically limited so it happens fast and completes at the early stage where the islands have smaller size and prefer an asymmetric transition path; when $r$ is small, the shape transition is dominated by kinetic limitations and many islands have to wait to transform until they grow to a larger size, in which case the symmetric transition path is energetically preferred. On the other hand, the deposition rate $\Phi$ indicates the rate of change of the total amount of adatoms in the system and hence controls the growth speed of the whole island population. For a fixed $r$, a larger $\Phi$ will force the island population grow faster so that more islands do not have enough time for transition at the early stage and will go through the symmetric transition at a later stage, as shown along rows in Table.~\ref{table}. From the table, it is also interesting to observe that the asymmetric island fraction in the multifaceted domes is typically higher than that in the transitional domes, which results from the higher energy barrier for the asymmetric transition P-H$_1$-D$_1$ than the asymmetric transition D$_1$-H$_m$-D$_m$.

\section{Discussion} \label{section:discussion}
\subsection{Creation of uniform arrays of quantum dots} \label{subsection:uniform}
Since it is still a challenge to control the shape, size and hence the properties of nanostructures by utilizing the self-assembly of quantum dots, \cite{Barth2005Nature} many theories and methods have been developed and discussed in literature. For example, Ni \textit{et al.} \cite{Ni2005three} suggested utilizing the strength of elastic anisotropy and appropriate epitaxial orientation to obtain favorable surface morphologies of quantum dots; Li \textit{et al.} \cite{Li2007Thermodynamic} and Aqua \textit{et al.} \cite{Aqua2015pitpattern} studied the influence of the patterned substrate on the formation of quantum dots in order to create uniform arrays of quantum dots with desired localizations; Shchukin \textit{et al.} \cite{Shchukin2013manufacture} investigated a new kind of formation of 3D islands on a subcritical wetting layer caused by the deposition of a third nonwetting material on the subcritical layer. \par
Due to the complexity of the self-assembly process, the multimodal (bimodal) size distribution and the coexistence of different island types bring the difficulty of creating uniform arrays of quantum dots. Regarding the bimodal distribution, Jesson and Munt \cite{Jesson2004Suppression} presented a mean-field kinetic model incorporating the elastic interactions between islands to study the quantum dot coarsening process and found that the elastic interactions will suppress the coalescence of islands and result in the absence of the bimodal distribution. In Refs.~\onlinecite{Munt2004Manipulating,Munt2007}, they applied the mean-field Fokker-Planck model to study the coarsening of quantum dots with uniform shape which possess a minimum energy per unit volume as a function of island size. Due to the existence of the minimum energy per volume, they proposed a method to tune the distribution of islands to a narrow metastable distribution at a desired size. In this paper, we construct a mean-field Fokker-Planck reaction model, which does not include the effects of elastic interactions and the minimum energy per unit volume, to study the distribution evolution of different island types coupled with the shape transitions. Based on our FPR model, we provide a new mechanism to prohibit the coarsening of the islands and produce a unimodal distributed array of islands by increasing the deposition flux rate or the system temperature. \par

\subsection{Comparison to experiments and simulations} \label{subsection:experiment}
The shape transition between pyramids and domes and the resulting bimodal distribution have been well studied in experiments and theories. \cite{Ross1998Coarsening,Rastelli2005,Vine2005} Also, the asymmetric shape transition during island growth has been observed and studied in experiment, \cite{ross1999transition} simulation \cite{lam2010kinetic} and theories. \cite{Spencer2013,Wei20160262} However, the evolution of the shape and size distribution of asymmetric islands is still not well-investigated due to the difficulty of calculating the energy barrier of asymmetric shape transitions. One important feature of our FPR model is to couple this asymmetric shape transition process with the island growth process by introducing a reaction term into the Fokker-Planck equation. Our model predicts the coexistence of different island types at the same size around the critical shape transition volume, and also describes the development of bimodal (and multimodal) size distributions depending on the growth conditions, which are consistent with the experiments. \cite{Ross1998Coarsening,Rastelli2005} In addition, we find that the asymmetric islands which appear as a metastable state of the shape transition have relatively smaller distribution and shorter lifetime during the evolution compared to the symmetric stable states, which results from the higher first barrier (P-H$_1$ or D$_1$-H$_m$) and lower second barrier (H$_1$-D$_1$ or H$_m$-D$_m$). \cite{Wei20160262} This explains the difficulty of observing asymmetric transitional states in experiments. Moreover, our results in Sec.~\ref{subsection:asymmetry} display similar behavior as in Ref.~\onlinecite{lam2010kinetic} regarding the competition between the kinetics of the shape transition and island growth. We find the shape transition is delayed to a larger size when the transition process is dominated by kinetics, which is indicated by the small transition/growth rate ratio $r$ in our model and corresponds to the case of low temperature in Ref.~\onlinecite{lam2010kinetic}. In addition, this delay of the shape transition turns out to affect the fraction of asymmetric transition islands within the whole island population. Since the asymmetric island shape is no longer a metastable state at larger sizes, more islands who grow to a larger size due to the delay will undergo the symmetric shape transition path and the fraction of asymmetric transition islands will decrease as shown in Table.~\ref{table}. Moreover, we find that the change in the asymmetric transition fraction will be enhanced if deposition flux of materials (large $\Phi$) provides sufficient adatoms in the reservoir for the island growth.

\subsection{Future generalization of FPR model} \label{subsection:generalization}
In this article, we present a Fokker-Planck reaction model for describing the coupled growth-transition process of quantum dots, which introduces into a system of Fokker-Planck equations reaction terms calculated from the lowest-barrier transition path on the energy potential surface of a reduced dimensional space as discussed at end of Sec.~\ref{subsection:reaction}. In this reduced space, we regard the island volume (size dependence) as the island growth coordinate and the length of transitional facets on the island (shape dependence) as shape transition coordinate. For the island growth, we calculate the energy of the equilibrium states as a function of island size. For the shape transition, at each island volume we obtain an energy surface as a function of the lengths of transitional facets and find a lowest-barrier transition path between different island types (equilibrium states) on the energy surface. This model captures the size and shape dependence of island energy and enables us to consider the evolution of the shape distribution for stable and metastable island types, whereas it excludes the intermediate (nonequilibrium) island states during the process. However, it is possible to give a full description of the growth-transition process for all possible faceted island states by investigating the Fokker-Planck (or Smoluchowski) model with our energy in Sec.~\ref{subsection:energy} in the high-dimensional space of islands composed of arbitrary facet lengths. The island size and shape dependence is embedded in this high-dimensional space and it would enable us to consider the island distribution as a function of all island facets which includes all faceted island shapes instead of only equilibrium states in the reduced space. \par
It is also worth noting that a quasi-steady-state assumption \cite{Munt2007,Atwater1990quasisteady} is applied in our model that the adatom concentration on the substrate has reached a quasi-steady-state and all the deposited adatoms are assumed to incorporate in the growing islands. With this assumption, we focus on the evolution of existing islands without considering the change of free adatoms in the reservoir and the formation of new islands in our system. Our FPR model could also be generalized to describe the change of free adatoms, the distribution and barrierless formation of prepyramids (small unfaceted islands) and the first-order shape transition between prepyramids and pyramids, \cite{Tersoff2002barrierless} which happen at the very early stage of the epitaxial growth of quantum dots.

\section{CONCLUSION} \label{section:summary}
We have developed a Fokker-Planck reaction model by introducing a reaction term into a mean-field Fokker-Planck model which enables us to describe the dynamics of shape transitions between different island types (eg, pyramid and dome). The resulting FPR model thus lets us describe how the shape and size distribution of an array of islands will evolve during the growth process. The reaction rate terms in the FPR model are determined from the results of the size-dependent lowest-barrier shape transition path between island \mbox{types} as calculated from a 2D energy model containing elastic energy and surface energy for faceted strained islands. Through simulations with different growth parameters, we find the unimodal and multimodal evolution modes of the island shape distribution that are \mbox{mainly} dependent on the external deposition flux rate and the temperature. In general, large deposition rate and high temperature promote the unimodal evolution mode due to the higher mean-field chemical potential. In addition, we investigate the importance of asymmetric transition shapes on the evolution of the shape distribution by varying the shape transition rate parameter. The results show that the reaction rates do not change the evolution mode but do affect the fraction of asymmetric transition islands in the island population, with faster transition rates resulting in asymmetric transitions dominating the process.

\bibliographystyle{h-physrev}
\bibliography{myreference}

\end{document}